\documentclass{aa}  
\usepackage{txfonts}
\usepackage{amsmath,amstext}
\usepackage{graphicx}
\usepackage{epstopdf}
\usepackage{float}
\usepackage{array}
\usepackage{mathtools}
\usepackage{booktabs}
\usepackage{subfigure}
\usepackage{url}
\usepackage{helvet}
\usepackage{tabularx}
\usepackage{multirow}
\usepackage{natbib}
\usepackage[flushleft]{threeparttable}
\usepackage{lscape}
\usepackage{pdflscape}
\usepackage{longtable}
\usepackage{wasysym}
\usepackage{float}
\usepackage{relsize}
\usepackage{color}
\usepackage[breaklinks=true]{hyperref} 
\usepackage{breqn}
\usepackage{bm}

\hypersetup{colorlinks,breaklinks, linkcolor=blue,urlcolor=blue, anchorcolor=blue,citecolor=blue}

\def\kms{km s$^{-1}$}         
\def\ms{\hbox{m s$^{-1}$}}         
\def\gcm3{\hbox{g cm$^{-3}$}}       
\def\vsini{\hbox{$\upsilon \sin i_{\star}$}}      
\def\Msun{\hbox{$\mathrm{M}_{\astrosun}$}}             
\def\Rsun{\hbox{$\mathrm{R}_{\astrosun}$}}

\def\Mearth{\hbox{$\mathrm{M}_{\oplus}$}}
\def\Rearth{\hbox{$\mathrm{R}_{\oplus}$}}
\def\degr{\hbox{$^\circ$}}
\def\teff{T$_{\rm eff}$}
\def\logg{log~{\it g}}
\def\met{[Fe/H]}
\def\vmicro{$\upsilon_\mathrm{micro}$}
\def\vmacro{$\upsilon_\mathrm{macro}$}

\bibpunct{(}{)}{;}{a}{}{,} 

\begin{document}

   \title{Exoplanet characterisation in the longest known resonant chain: the K2-138 system seen by HARPS  \thanks{Table \ref{RVtable} is available in electronic form at the CDS via anonymous ftp to \url{cdsarc.u-strasbg.fr} (130.79.128.5)
or via \url{http://cdsweb.u-strasbg.fr/cgi-bin/qcat?J/A+A/}}}
   
\author{       T.~A.~Lopez\inst{\ref{LAM}}
       \and    S.~C.~C.~Barros\inst{\ref{IA}}
       \and    A.~Santerne\inst{\ref{LAM}}
       \and    M.~Deleuil\inst{\ref{LAM}} 
       \and    \\V.~Adibekyan\inst{\ref{IA}}
       \and    J.-M.~Almenara\inst{\ref{geneva}}
       \and    D.~J.~Armstrong\inst{\ref{warwick},\ref{CEH}}
       \and    B.~Brugger\inst{\ref{LAM},\ref{Corn}}
       \and    D.~Barrado\inst{\ref{CAB}}
       \and    D.~Bayliss\inst{\ref{warwick}}
       \and    I.~Boisse\inst{\ref{LAM}}
       \and    A.~S.~Bonomo\inst{\ref{Torino}}
       \and    F.~Bouchy\inst{\ref{geneva}}
       \and    D.~J.~A.~Brown\inst{\ref{warwick},\ref{CEH}}     
       \and    E.~Carli\inst{\ref{Glasgow}}  
       \and    O.~Demangeon\inst{\ref{IA}}
       \and    X.~Dumusque\inst{\ref{geneva}}       
       \and    R.~F.~D\'iaz\inst{\ref{UBA},\ref{CONICET},\ref{geneva}}
       \and    J.~P.~Faria\inst{\ref{IA},\ref{UPorto}}
       \and    P.~Figueira\inst{\ref{IA}}    
       \and    E.~Foxell\inst{\ref{warwick}}    
       \and    H.~Giles\inst{\ref{geneva}}
       \and    G.~H\'ebrard\inst{\ref{IAP},\ref{OHP}}
       \and    S.~Hojjatpanah\inst{\ref{IA},\ref{UPorto}}          
       \and    J.~Kirk\inst{\ref{CfA}} 
       \and    J.~Lillo-Box\inst{\ref{ESO},\ref{CAB}}
       \and    C.~Lovis\inst{\ref{geneva}}
       \and    O.~Mousis\inst{\ref{LAM}}        
       \and    H. ~J. ~da~Nóbrega\inst{\ref{UPorto},\ref{CAP}} 
       \and    L.~D.~Nielsen\inst{\ref{geneva}}
       \and    J.~J.~Neal\inst{\ref{IA},\ref{UPorto}} 
       \and    H.~P.~Osborn\inst{\ref{LAM}}
       \and    F.~Pepe\inst{\ref{geneva}}
       \and    D.~Pollacco\inst{\ref{warwick},\ref{CEH}}   
       \and    N.~C.~Santos\inst{\ref{IA},\ref{UPorto}}    
       \and    S.~G.~Sousa\inst{\ref{IA}}   
       \and    S.~Udry\inst{\ref{geneva}}
       \and    A.~Vigan\inst{\ref{LAM}}
       \and    P.~J.~Wheatley\inst{\ref{warwick},\ref{CEH}}
       }      

   \institute{
   		Aix Marseille Univ, CNRS, CNES, LAM, Marseille, 					France\label{LAM}
 	 	\and Instituto de Astrof\'isica e Ci\^{e}ncias do Espa\c co, Universidade 		do Porto, CAUP, Rua das Estrelas, 4150-762 Porto, Portugal\label{IA}
 		\and Observatoire Astronomique de l'Universit\'e de Gen\`eve, 51 			Chemin des Maillettes, 1290 Versoix, Switzerland\label{geneva} 	
 		\and Department of Physics, University of Warwick, Gibbet Hill Road, 			Coventry, CV4 7AL, UK\label{warwick} 		 	
  		\and Centre for Exoplanets and Habitability, University of Warwick, 			Gibbet Hill Road, Coventry CV4 7AL, UK\label{CEH}
		\and Cornell Center for Astrophysics and Planetary Sciences, Department of Astronomy, Cornell University, Ithaca, NY, USA \label{Corn}
		\and Depto. de Astrof\'isica, Centro de Astrobiolog\'ia (CSIC-INTA), 			ESAC campus 28692 Villanueva de la Ca\~nada (Madrid), 				Spain\label{CAB}
 		\and INAF -- Osservatorio Astrofisico di Torino, Strada Osservatorio 20, 		I-10025, Pino Torinese (TO), Italy\label{Torino}
  		\and University of Glasgow, Glasgow, G12 8QQ, Scotland\label{Glasgow}
		\and Universidad de Buenos Aires, Facultad de Ciencias Exactas y 		Naturales. Buenos Aires, Argentina\label{UBA}
		\and CONICET - Universidad de Buenos Aires. Instituto de 				Astronom\'ia y F\'isica del Espacio (IAFE). Buenos Aires, 					Argentina\label{CONICET}
		\and Departamento de F\'isica e Astronomia, Faculdade de Ciencias, 		Universidade do Porto, Rua Campo Alegre, 4169-007 Porto, 				Portugal\label{UPorto}	
		\and Institut d'Astrophysique de Paris, UMR7095 CNRS, Universite 		Pierre \& Marie Curie, 98bis boulevard Arago, 75014 Paris, 				France\label{IAP}
		\and Aix Marseille Univ, CNRS, OHP, Observatoire de Haute 			Provence, Saint Michel l'Observatoire, France\label{OHP}
  		 \and Center for Astrophysics | Harvard \& Smithsonian, 60 Garden Street, Cambridge, MA 02138, USA  \label{CfA}
  		 \and CAP – Centre for Applied Photonics, INESC TEC, Porto, Portugal \label{CAP}
 		\and European Southern Observatory (ESO), Alonso de Cordova 3107, 		Vitacura, Casilla 19001, Santiago de Chile, Chile\label{ESO}
       	}

   \date{Received July 8, 2019; accepted September 20, 2019}
 
  \abstract{The detection of low-mass transiting exoplanets in multiple systems brings new constraints to planetary formation and evolution processes and challenges the current planet formation theories. Nevertheless, only a mere fraction of the small planets detected by \textit{Kepler} and \textit{K2} have precise mass measurements, which are mandatory to constrain their composition. We aim to characterise the planets that orbit the relatively bright star K2-138. This system is dynamically particular as it presents the longest chain known to date of planets close to the 3:2 resonance. We obtained 215 HARPS spectra from which we derived the radial-velocity variations of K2-138. Via a joint Bayesian analysis of both the \textit{K2} photometry and HARPS radial-velocities (RVs), we constrained the parameters of the six planets in orbit. The masses of the four inner planets, from b to e, are $3.1$, $6.3$, $7.9$, and $13.0$ \Mearth\ with a precision of 34\%, 20\%, 18\%, and 15\%, respectively. The bulk densities are $4.9$, $2.8$, $3.2$, and $1.8$ \gcm3, ranging from Earth  to Neptune-like values. For planets f and g, we report upper limits. Finally, we predict transit timing variations of the order two to six minutes from the masses derived. Given its peculiar dynamics, K2-138 is an ideal target for transit timing variation (TTV) measurements from space with the upcoming CHaracterizing ExOPlanet Satellite (CHEOPS) to study this highly-packed system and compare TTV and RV masses.}
 

   \maketitle
%

\section{Introduction}
Precise knowledge of both the mass and radius of planets is necessary before initiating more advanced studies. This is particularly true for internal structure modelling which, beyond the degeneracies inherent to this type of investigation, requires a precise determination of the planetary bulk density \citep{2007ApJ...669.1279S, 2017ApJ...850...93B}. This is also true for atmospheric studies which rely on mass measurements to derive the scale height, which is an essential parameter to interpret observations in transmission spectroscopy (e.g. \citealt{2009ASSP...10..123S}). Few low-mass planets have both a precise radius and mass. One of the reasons is that the \textit{Kepler} mission focussed on faint stars (V$>13$), prohibiting subsequent Doppler monitoring from the ground for most of them. 

Understanding the compositions of small planets allows us to put strong constraints on their formation path, and even helps to understand the formation of the solar system (e.g. \citealt{2018MNRAS.479L..81R}). For example, \textit{Kepler} provides evidence for a bi-modality in the radius distribution of small planets, with a gap around 1.6 \Rearth\ \citep{2015ApJ...801...41R, 2017AJ....154..109F, 2018AJ....156..264F}. Moreover, there are several examples of planets with a similar radius but radically different compositions: \textit{Kepler}-11f and \textit{Kepler}-10c have a radius of $2.61\pm0.25$ \Rearth\ and $2.35\pm0.05$ \Rearth\, respectively, whereas the first is gaseous and the second is rocky \citep{2011Natur.470...53L, 2014ApJ...789..154D}. This has also been observed for planets within the same system \citep{2019NatAs...3..416B} and planets with sizes in the range of 1.5 \Rearth\ to 4 \Rearth\ can have compositions ranging from gaseous (dominated by H-He) to rocky (dominated by iron and silicates), and even to icy (water-worlds with a large amount of volatiles) \citep{2004Icar..169..499L, 2011ApJ...738...59R, 2015ApJ...801...41R}. The transition between rocky and gaseous planets is not thoroughly understood and more well-characterised low-mass planets are required. The \textit{K2} mission provides us with a large number of small planet candidates around relatively bright stars, allowing precise spectroscopic follow-up from the ground to determine their masses (e.g. \citealt{2017A&A...604A..19O}, \citealt{2017A&A...608A..25B}, \citealt{2018NatAs...2..393S}, \citealt{2018A&A...620A..77L}).

Among these, some are multiple systems of small transiting planets. They are really valuable as they highlight differences in planetary formation and evolution within the same environment, allowing us to perform comparative studies \citep{2019NatAs...3..416B}. These systems are often compact and coplanar \citep{2011Natur.470...53L, 2014ApJ...790..146F, 2015ARA&A..53..409W} and bring further constraints to the evolution and migration models. Indeed, they are believed to have migrated to their current position rather than formed in-situ (e.g. \citealt{2017A&A...604A...1O}). In addition, some of these tightly-packed systems present resonances, which are a consequence of inward migration when drifting planets pile-up in or near mean-motion resonances, thus creating chains of resonances \citep{2008MNRAS.384..663R, 2014ApJ...790..146F, 2017A&A...604A...1O}.

This is the case of the K2-138 planetary system. K2-138 is a moderately bright star (V $=12.2$) observed in the campaign 12 of the \textit{K2} mission. Four planets were initially identified by citizen scientists involved in the Exoplanet Explorers project\footnote{\url{www.exoplanetexplorers.org}}. A detailed analysis of the light curve revealed the presence of potentially six planets. The first five form a compact inner system with radii ranging from 1.6 \Rearth\ to 3.3 \Rearth\ and periods between 2.35 days and 12.76 days \citep{2018AJ....155...57C}. For the outer planet, only two transits are available in the \textit{K2} data, indicating a potential period of 42 days. Additional Spitzer observations confirm its planetary nature (Hardegree-Ullman et al., in preparation). The five inner planets form a chain of near first-order 3:2 mean-motion resonance, and all the planets lie just outside these resonances. This is the longest known chain to date \citep{2014ApJ...790..146F}. The five inner planets also are in a chain of three-body Laplace resonances \citep{2018AJ....155...57C}, which is similar to the TRAPPIST-1 system \citep{2017NatAs...1E.129L}. 

The brightness of K2-138 makes it an interesting target for ground-based spectroscopic follow-up. The precise mass measurement of these small planets is within reach of an instrument like the High Accuracy Radial velocity Planet Searcher (HARPS) on a 3.6 m telescope. In this paper, we present radial velocity (RV) observations acquired in 2017 and 2018, allowing us to constrain precisely the masses of three of the six planets in the system, and to put strong constraints on the others. We used these measurements to predict transit timing variations (TTVs) for the purpose of observing them with the CHaracterizing ExOPlanet Satellite (CHEOPS). The observations and data reduction are described in Sect. \ref{sect2}. The spectral analysis, activity characterisation, and the joint analysis of the radial velocities and photometry are reported in Sect \ref{sect3}. Finally, we discuss the validity of the results and the magnitude of transit timing variations in Sect. \ref{sect4}. 

\section{Observations and data reduction \label{sect2}}

\subsection{K2 Photometry \label{photometry}}

\begin{figure}[htbp!]
	\centering
	    \includegraphics[width=0.5\textwidth]{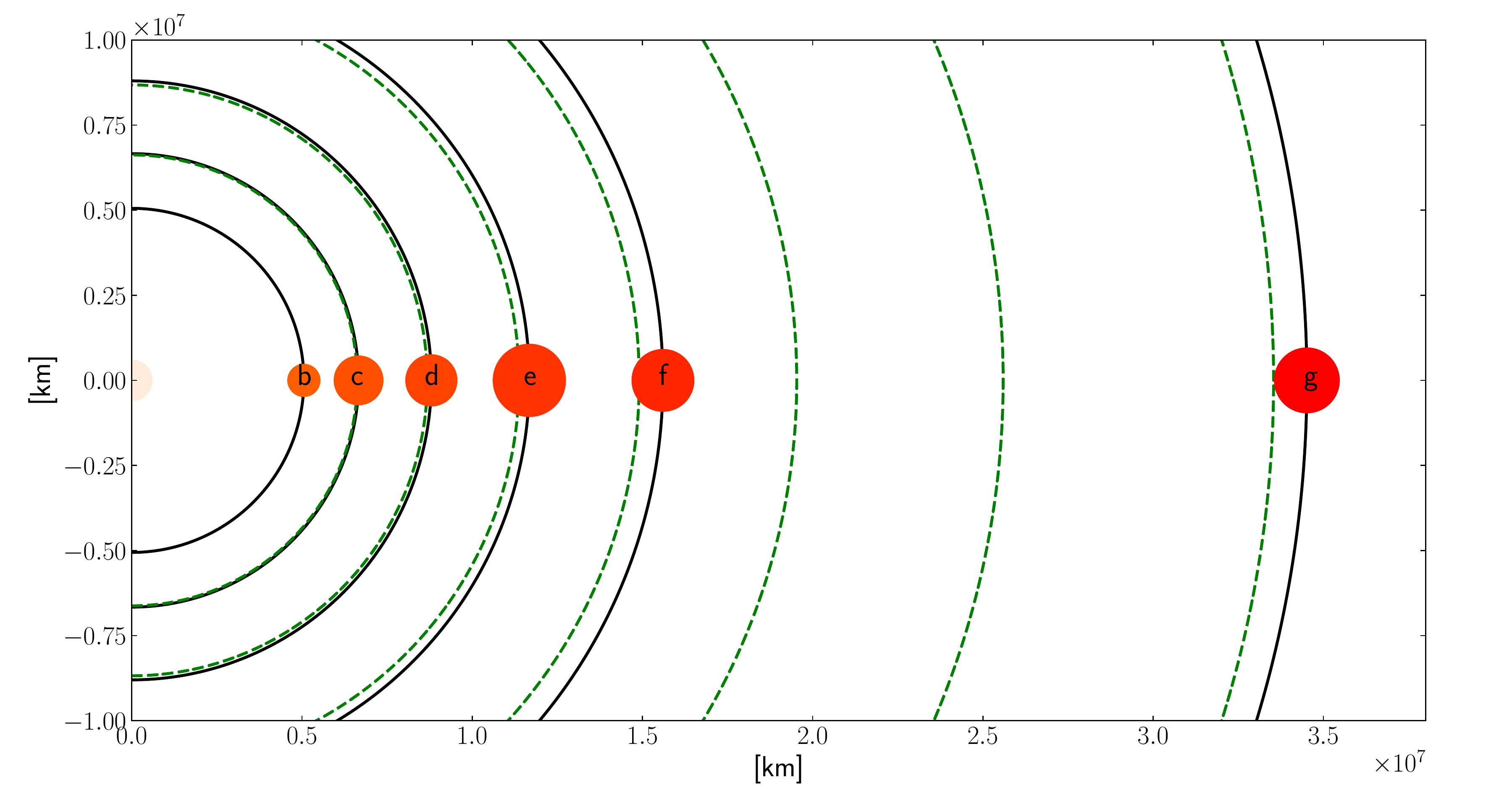} 
	    \caption{\label{system} K2-138 planetary system. All distances are drawn to scale. Planetary radii are enlarged by a factor 50 for readability. Orbits, assumed circular, are shown in black whereas the location of the 3:2 resonances are shown in dashed green.}
	\end{figure}

K2-138 was observed for 79 days during the campaign 12 of the \textit{K2} mission \citep{2014PASP..126..398H} between December 15, 2016 and March 04, 2017 in long cadence mode. On February 01, 2017 the spacecraft entered safe mode resulting in a 5.3 days gap. Five distinct transiting signals were identified and analysed as described in \citet{2018AJ....155...57C} leading to the discovery of five sub-Neptune planets with sizes ranging from 1.6 \Rearth\ to 3.3 \Rearth\ in a chain of near 3:2 resonances. Additionally, a potential sixth planet was identified with a 42 days period. Hardegree-Ullman et al. (in preparation) were able to confirm it using observations from the Spitzer Space Telescope. False-positive scenarios are discussed and ruled out in \citet{2018AJ....155...57C}. A sketch of the system is shown in Fig. \ref{system} along with the planet distances to the resonances.

Our group independently discovered K2-138 using a new transit search routine that was optimised for multi-planetary systems. The code is based on a Python re-implementation\footnote{\url{https://github.com/dfm/python-bls}} of the Box-fitting least squares (BLS) \citep{Kovacs} algorithm. To search for multi-planetary systems, the code repeats the BLS transit search several times per light curve after removing the model of transits found in the previous BLS search. For each campaign we usually apply the BLS three times to all the light curves reduced with the POLAR pipeline. The candidates that pass the vetting procedure were re-analysed with our transit searching code with a higher number of BLS calls to allow the discovery of more planets in the system. For K2-138 the first search revealed two candidates with periods 5.4 days and 8.3 days. Other transits were also clearly seen in the light curve by eye.  For the second search we performed the BLS six times and found two clearer candidates with periods 3.6 days and 2.35 days and a possible candidate with a period of 12.8 days. As a very interesting planetary transiting system, we started the RV follow-up with HARPS early on after its identification by our group.

We used the Planet candidates from OptimaL Aperture Reduction (\texttt{POLAR}) pipeline \citep{2016A&A...594A.100B} to reduce the pixel data and extract the light curve. The \texttt{POLAR} pipeline is summarised as follows: We downloaded the calibrated pixel data (pixel files) from the Mikulski Archive for Space Telescopes (MAST)\footnote{\url{http://archive.stsci.edu/kepler/data_search/search.php}} and performed a photometric extraction using the adapted CoRoT imagette pipeline from \citet{2014A&A...569A..74B}, which uses an optimal aperture following the point spread function. The centroid positions were then computed using the Modified Moment Method \citep{1989AJ.....97.1227S} to correct the flux-position systematics. This correction was performed using a self-flat-fielding method similar to \citet{2014PASP..126..948V}. The first transit (planet d) was removed due to the significantly higher noise level at the very beginning of the campaign. We also used the open source freely available\footnote{\url{https://rodluger.github.io/everest/using_everest.html}} EPIC Variability Extraction and Removal for Exoplanet Science Targets (\texttt{EVEREST}) pipeline \citep{2016AJ....152..100L, 2018AJ....156...99L} on the transit-filtered light curve. \texttt{EVEREST} rely on an extension of the pixel level decorrelation used on Spitzer data \citep{2015ApJ...805..132D}. We used both the \texttt{POLAR} and \texttt{EVEREST} light curves for a complete combined analysis with the radial velocities. We report the results from the  \texttt{EVEREST} reduction as it has slightly better K2 residuals and allows for a better convergence of the analysis. Nevertheless, they are consistent with the \texttt{POLAR} ones. Figure \ref{k2lc} shows the extracted light curve from \texttt{EVEREST}. We also corrected the K2 light curve for spot and faculae modulation: the light curve was first detrended using a second order polynomial regression to remove low-frequency variations, outliers were removed with a sigma-clipping and the activity was filtered as described in Sect.~\ref{analysis}. Finally, we performed the transit fitting jointly with the radial velocity analysis as described in details in Sect.~\ref{analysis}.

\subsection{HARPS radial-velocity follow-up}

We collected 215 spectra of K2-138 over 79 nights between September 25, 2017 and September 04, 2018 with the HARPS spectrograph (R $\sim 115,000$) mounted on the 3.6 m Telescope at ESO La Silla Observatory \citep{2003Msngr.114...20M}. These observations were conducted as part of the ESO-K2 large programme (ID 198.C-0169) with the aim of detecting the masses of the five transiting planets known at this time. We used an exposure time of 1800s, giving a mean signal-to-noise ratio (S/N) of 28.5 per pixel at 550 nm. The spectra were reduced using the HARPS pipeline. Radial velocities (RVs) were derived by cross-correlating the spectra with a numerical template of a G2V star \citep{1996A&AS..119..373B, 2002Msngr.110....9P} and the photon noise was estimated as described in \citet{2001A&A...374..733B}. The average photon noise is 3.92 \ms. Finally, the full width half maximum (FWHM) and the bisector inverse slope (BIS) of the cross-correlation functions were measured and their uncertainties computed as described in \citet{2015MNRAS.451.2337S}. We rejected 21 measurements that were affected by Moon contamination, showing significant anomalies in their RV and FWHM. The remaining 194 measurements were used for the analysis described in Sect. \ref{analysis} and are reported in Table \ref{RVtable}. We also measured indices sensitive to stellar activity: H$_{\alpha}$, Na~{\sc i} D and Ca~{\sc ii} from which we derived S$_{\rm MW}$. The periodograms associated to the RVs and different indices are given in Fig. \ref{LSplot}.

\subsection{Gaia astrometry \& contamination \label{gaia}}  

We used the \textit{Gaia} Data Release 2 \citep{2018A&A...616A...1G} parallax, taking into account the correction from \citet{2019MNRAS.487.3568S}, to derive the stellar distance. K2-138 has a corrected parallax of $4.962 \pm 0.065$ mas, leading to a distance of $201.66 \pm 6.38$ pc. Table \ref{stellarparam1} gives the astrometric properties of K2-138. \citet{2018A&A...616A...8A} derived the effective temperature from the three-band photometry and the parallax, giving $\rm T_{eff} = 5110_{-53}^{+204}$ K. They also used the extinction $A_G = 0.0610_{-0.0125}^{+0.2050}$ and the reddening $E(BP-RP) = 0.0308_{-0.0056}^{+0.1006}$ to determine the luminosity $L = 0.517_{-0.07}^{+0.07}~L_{\odot}$ and radius $R_s = 0.92_{-0.07}^{+0.02}~R_{\odot}$. The stellar parameters from \textit{Gaia} DR2 are consistent with the distance estimate, effective temperature and stellar radius which are derived in the joint Bayesian analysis in section \ref{analysis} and via the spectral analysis. In complement to the planetary transit validation in \citet{2018AJ....155...57C}, we also checked the presence of contaminants resolved by \textit{Gaia}. There is only one source already reported in \citet{2018AJ....155...57C}, within 14$^{\prime\prime}$ with a G mag of 20.1, giving $\Delta m = 8.1$. The contamination is therefore negligible, at most at the level of $10^{-3}$, assuming the contaminant falls within the K2 aperture.

	\begin{table}
  \begin{threeparttable}
    \caption{\label{stellarparam1}Stellar properties of K2-138.}
    \small
     \begin{tabular}{lc}
        \toprule
        Parameter & Value and uncertainty \\
        \midrule
        \hline
		\multicolumn{2}{l}{\textit{Astrometry}}	\\
		\\
		\textit{K2} campaign & 12$^a$ \\
		EPIC & 245950175$^a$\\
		2MASS ID & J23154776$-$1050590$^b$\\
		Gaia ID & DR2 2413596935442139520$^c$ \\
		RA (J2000) &  23:15:47.77$^c$ \\
		Dec (J2000) & $-$10:50:58.90$^c$ \\ 
		$\mu_{RA}$ [mas/yr] & $-1.02 \pm 0.09^c$ \\
		$\mu_{DEC}$ [mas/yr] & $-10.52 \pm 0.09^c$\\
		Parallax [mas] & $4.962 \pm 0.065^d$ \\
		Distance [pc] & $201.66 \pm 6.38^d$ \\
		\\
		\hline
		\multicolumn{2}{l}{\textit{Photometric magnitudes}}	\\
		\\
		Kepler Kp & $12.069^a$ \\ 
		Gaia G & $12.0252\pm0.0004^c$ \\ 
		Gaia BP & $12.4634\pm0.0022^c$ \\ 
		Gaia RP & $11.4524\pm0.0015^c$ \\ 
		Johnson B & $13.063\pm0.051^e$ \\
		Johnson V & $12.217\pm0.035^e$ \\
		Sloan g$^{\prime}$ & $12.669\pm0.262^e$\\
		Sloan r$^{\prime}$ & $11.962\pm0.038^e$\\
		Sloan i$^{\prime}$ & $11.809\pm0.209^e$\\
		2-MASS J & $10.756\pm0.021^b$\\
		2-MASS H & $10.384\pm0.021^b$ \\
		2-MASS Ks & $10.305\pm0.021^b$ \\
		WISE W1 & $10.274\pm0.023^f$ \\
		WISE W2 &$10.332\pm0.019^f$ \\
		WISE W3 & $10.279\pm0.077^f$\\
		\\
		\hline
		\multicolumn{2}{l}{\textit{Stellar parameters (adopted)}}	\\
		\\
		Effective temperature $T_{\rm eff}$ [K] & $5350 \pm 80^g$ \\ 
		Surface gravity $\log g$ [cgs] & $4.52 \pm 0.15^g$ \\ 
		Iron abundance \met\ [dex] & $0.14 \pm 0.10^g$  \\
		Metallicity [M/H] [dex] & $0.15\pm0.04^g$  \\
		Rotational velocity \vsini ~ [\kms] & $2.5 \pm 1.0^g$  \\
		Macroturbulence \vmacro ~ [\kms] & $1.9 \pm 0.1^g$  \\
		Microturbulence \vmicro ~ [\kms] & $0.90 \pm 0.10^g$  \\
		Stellar age (gyrochronology) [Gyr] & $2.3^{_{+0.44}}_{^{-0.36}}~^g$ \\
		Spectral type & G8\\	
        \bottomrule
     \end{tabular}
    \begin{tablenotes}
      \small
      \item a. EXOFOP-K2: \url{https://exofop.ipac.caltech.edu/k2/}
      \item b. Two Micron All Sky Survey (2MASS)
      \item c. \textit{Gaia} DR2
      \item d. Schoenrich, McMillan \& Eyer (2019). Distances for the Gaia RV set with corrected parallaxes (Version 1.0): \url{http://doi.org/10.5281/zenodo.2557803}
      \item e. AAVSO Photometric All-Sky Survey (APASS)
      \item f. AllWISE
      \item g. Spectral analysis (Sect. \ref{companion_star})
    \end{tablenotes}
  \end{threeparttable}
\end{table}

\section{Data analysis and results \label{sect3}}

	\begin{figure*}[htbp!]
	\centering
	    \includegraphics[width=\textwidth]{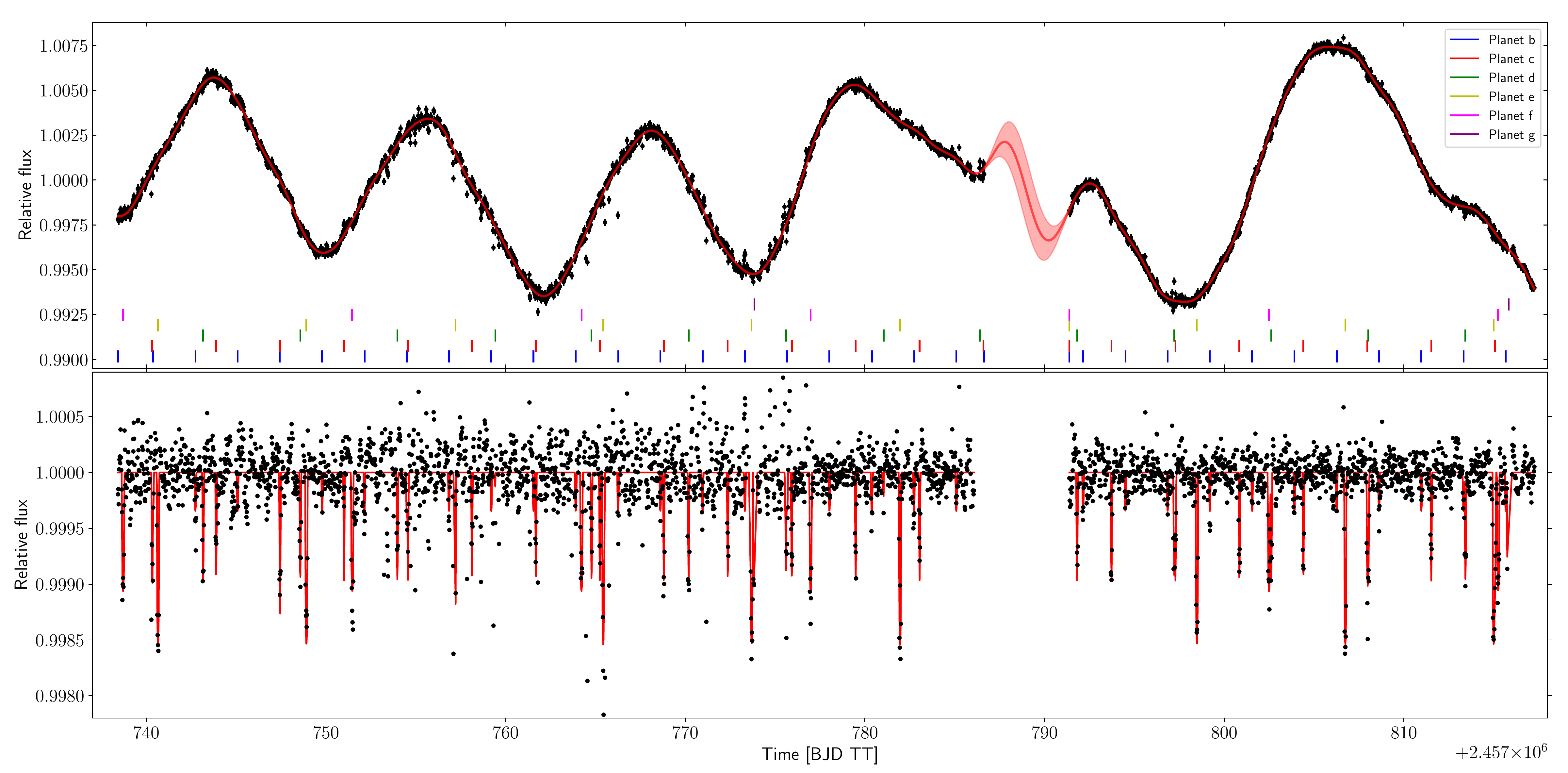} 
	    \caption{\label{k2lc} K2-138 light curve and associated models. \textit{Top:} Extracted \textit{K2} light curve, using the EVEREST pipeline, of K2-138 with in red the Gaussian process regression trained on the out-of-transit light curve. The positions of transits are marked with coloured lines at the bottom of the top figure. \textit{Bottom:} Light curve after subtraction of the Gaussian process fit. The best six-planet photometric model of the transits is superimposed in red.}
	\end{figure*}

	\begin{figure*}[htbp!]
	\centering
	    \includegraphics[width=\textwidth]{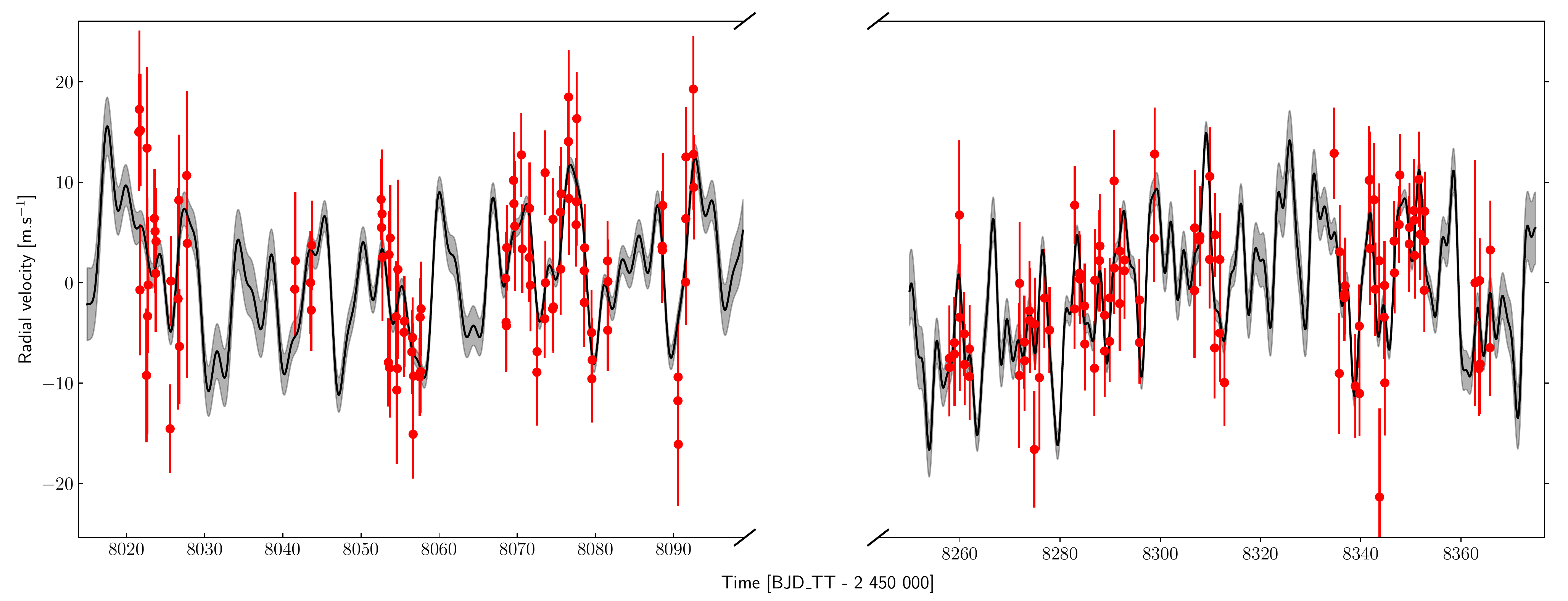} 
	    \caption{\label{rv} HARPS radial velocity time series obtained in two seasons (in red). The six-planets Keplerian model and the Gaussian process regression are shown in black with a 68.3\% confidence interval in grey.}
	\end{figure*}

\subsection{Spectral analysis of the host star \label{companion_star}}

The spectral analysis of the host star was performed to derive its fundamental parameters and the chemical composition of the photosphere. We used the HARPS spectra corrected from systemic velocity and planetary reflex-motion. We removed the spectra with a S/N lower than ten in order 47 (550 nm) and the ones contaminated by the Moon (S/N above 1.0 in fibre B). We then co-added the spectra in a single 1D spectrum. Then, the spectral analysis was performed using two sets of tools.

Firstly, we proceeded as described in \citet{2012A&A...538A.145D, 2014A&A...564A..56D}. The single 1D spectrum was carefully normalised. The spectral analysis was performed using the Versatile Wavelength Analysis (\texttt{VWA}) package \citep{2010A&A...519A..51B}. The spectral parameters were determined by fitting the Iron lines until the derived Fe~{\sc i} and Fe~{\sc ii} abundances minimised the correlation with both the equivalent width and the excitation potentials. The surface gravity was determined from the pressure-sensitive lines Mg~{\sc i} b, Na~{\sc i} D and the calcium lines at 612.2 nm, 616.2 nm and 643.9 nm. We found $\log g = 4.52\pm0.15$ [cgs], $T_{\rm eff}= 5350 \pm 80$ K, \vmicro $ = 0.90\pm0.10$ \kms , \vmacro $ = 1.9\pm0.1$ \kms , \vsini $=2.5\pm1.0$ \kms, [M/H]$=0.15\pm0.04$, and \met\ $= 0.14\pm0.10$. The abundances of elements with isolated lines were derived. They are shown in Table \ref{abundance}. For the Lithium, we do not find any significant {Li \sc i} line. We computed the stellar age $t = 2.3^{_{+0.44}}_{^{-0.36}}$ Gyr using the calibration with the chromospheric emission \citep{1993PhDT.........3D, 2004ApJS..152..261W}. The age derived from the joint analysis (Sect \ref{analysis}) is $2.8^{_{+3.8}}_{^{-1.7}}$ Gyr. Both are consistent. The final stellar parameters are given in Table \ref{stellarparam1}. They are all fully consistent with those obtained within the joint analysis with \texttt{PASTIS}, as described in Sect \ref{analysis}. They are also consistent with the results in \citet{2018AJ....155...57C}. 

As an independent check of the stellar parameters, we employed the same tools as used for stars in the SWEET-Cat catalogue \citep{2013A&A...556A.150S}. The method is in essence the same as in the first part (equivalent widths of Fe~{\sc i} and Fe~{\sc ii} lines, and iron excitation and ionisation equilibrium). Local thermodynamical equilibrium is assumed in the framework of the 2014 version of the code \texttt{MOOG}  \citep{1974ApJ...189..493S}, together with the grid of Kurucz \texttt{ATLAS9} plane-parallel model atmospheres \citep{1993yCat.6039....0K} and the equivalent width measurements of the iron lines from the code \texttt{ARES} \citep{2015A&A...577A..67S}. The derived values are $\log g = 4.327\pm0.069$ [cgs], $T_{\rm eff}= 5277 \pm 37$ K, \vmicro $ = 0.838\pm0.060$ \kms\, and \met\ $= 0.069\pm0.024$. The $\log g$ value, known to be systematically biased, was corrected using a calibration based on asteroseismic targets \citep{2014A&A...572A..95M}. 
The corrected value is $\log g = 4.51 \pm 0.07$ [cgs]. All these values agree with the previous ones, within uncertainties.

\subsection{Activity \& stellar rotation \label{rotation}}

\begin{figure}[htbp!]
	    \includegraphics[width=0.5\textwidth]{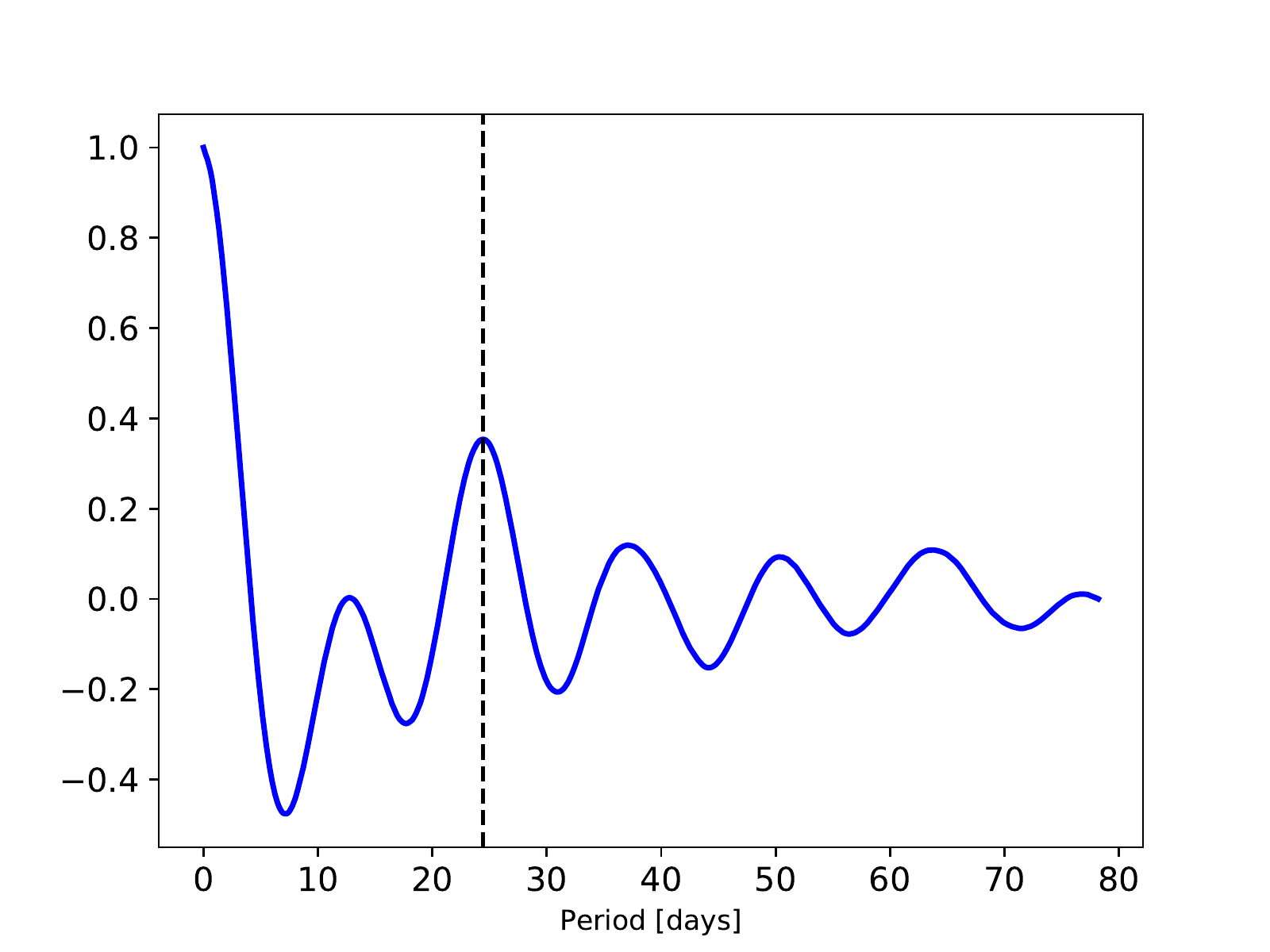}
	    \caption{\label{acf}Autocorrelation function of K2-138 light curve. The dashed line in black shows the dominant period, at 24.4 days, corresponding to the stellar rotation. It was computed on the smoothed ACF, following recommendations from \citet{2013MNRAS.432.1203M}.}
	\end{figure}

The stellar rotation period was first derived using the \textit{K2} light curve. We computed the autocorrelation function (ACF) (Fig. \ref{acf}) as described in \citet{2013MNRAS.432.1203M}. It shows a dominant periodicity at $24.44\pm2.35$ d. Then, we trained a Gaussian process (GP) with a quasi-periodic kernel (Eq. \ref{kernelrv} in Sect. \ref{analysis}) on the six hours-binned, transit-filtered light curve. We used a uniform prior on the GP period, between 15 and 40 days. All the priors are listed in Table \ref{MCMCprior}. We found a period of $24.7^{_{+0.3}}_{^{-2.2}}$ d. 

We also used the spectroscopic data to constrain the stellar rotation. The BIS periodogram (Fig. \ref{LSplot}) shows a peak with a false alarm probability (FAP) below 10 \% at around 12.5 d, which is close to half the rotation period. The H$_{\alpha}$ index and FWHM periodogram (Fig. \ref{LSplot}) show a significant peak at 25 days of period, below 0.1\% FAP. In contrast, there is no clear signal in the S$_{\rm MW}$ time series. We also found a decreasing linear drift in the FWHM, H$_{\alpha}$ and S$_{\rm MW}$ times series (Fig. \ref{LSplot}). This may be indicative of a magnetic cycle \citep{2011ESS.....2.0202L}. We then trained a GP on the FWHM, BIS,  H$_{\alpha}$, Na~{\sc i} D and S$_{\rm MW}$. The priors and posteriors of all the hyperparameters are reported in Table \ref{MCMCprior}. We obtained loose constraints of respectively $25.8^{_{+7.7}}_{^{-12}}$ d, $22 \pm 13$ d, $26.3^{_{+5.7}}_{^{-2.8}}$ d, $22 \pm 12$ d, and $23 \pm 12$ d.

Then, we derived the stellar rotation period from the $\log R'_{HK}$ index and the $\rm B-V$ colour as described in \citet{2008ApJ...687.1264M}. We computed the $\log R'_{HK}$ from the S$_{\rm MW}$ measurements and the APASS $\rm B-V$ colour \citep{2011AAS...21812601H}, following the calibrations from \cite{1984ApJ...287..769N} and the correction from \citet{2011arXiv1107.5325L} assuming an Iron abundance of $0.14\pm0.10$ and a $\rm B-V = 0.839\pm0.062$ corrected from the extinction. This led to an averaged value of $\log R'_{HK} = -4.76 \pm 0.04$. We deduced from it the Rossby number $R_0 = 1.51 \pm 0.11$ \citep{2008ApJ...687.1264M}. From the $\rm B-V$ colour, we derived the convective turnover time $\tau_c = 20.6 \pm 1.7$ d \citep{1984ApJ...287..769N}, and using the relation $P_{\rm rot} = R_0 \times \tau_c$, we derived $P_{\rm rot} = 31 \pm 4$ d. This value is compatible, within errors, with the ones derived above.

Finally, we derived the projected rotational velocity of K2-138 using calibrations from the FWHM. We obtained \vsini$ = 2.2 \pm 1.2$ \kms\ which we combined with the stellar radius from \textit{Gaia} DR2 (Sect. \ref{gaia}) to derive an upper limit on the stellar rotation period $P_{\rm rot} = 21.3 \pm 3.2$ d, assuming $i = \frac{\pi}{2}$.

In the end, as most of the indicators tend to show a stellar rotation period at 25 days, we modelled the activity-induced RVs using a Gaussian process regression as described in the next section with $P_{\rm rot} = 24.7\pm2.2$ d as prior of the stellar rotation period. This prior is drawn from the symmetrised value obtained with the GP learning on the light curve alone, taking a conservative width.

\subsection{\texttt{PASTIS} analysis \label{analysis}}

	\begin{figure*}[htbp!]
	\centering
	    \includegraphics[width=1.\textwidth]{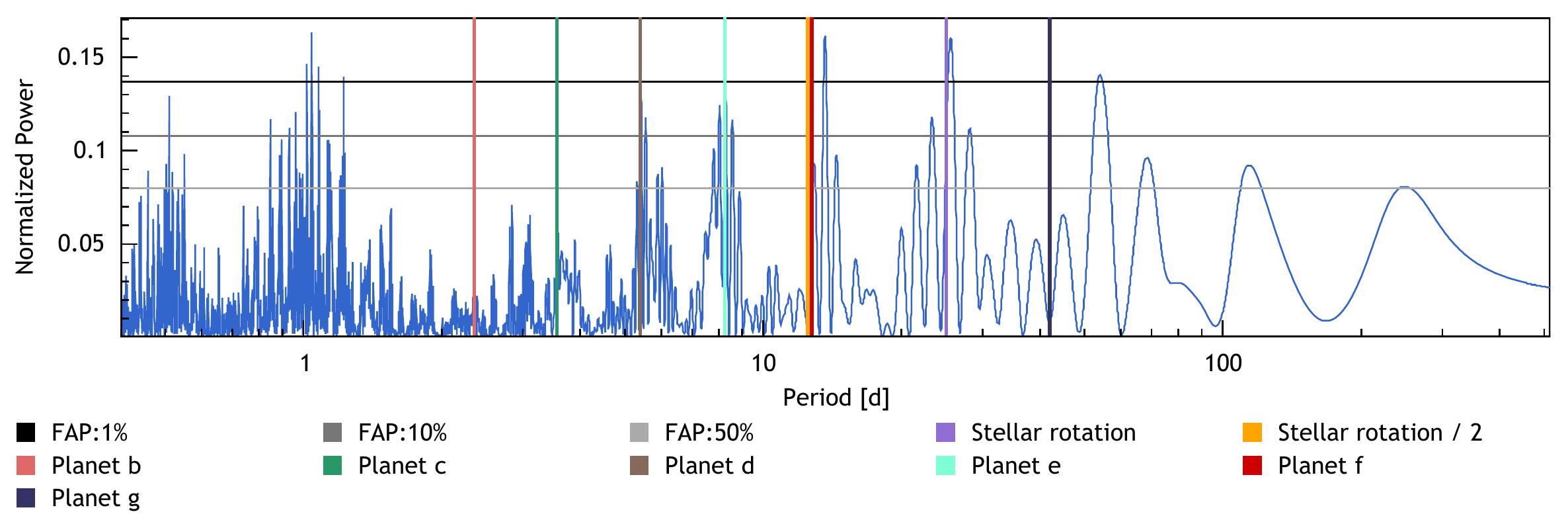}
		\caption{\label{rv-periodo} Lomb-Scargle periodogram of HARPS radial velocity. The peak position marked by the purple line corresponds to a period of $25$ days (stellar rotation). The orbital periods of the six planets are shown with coloured vertical lines. False alarm probability levels are shown with horizontal lines.}
	\end{figure*}

	\begin{figure}[htbp!]
	\centering
	    \includegraphics[scale=0.55]{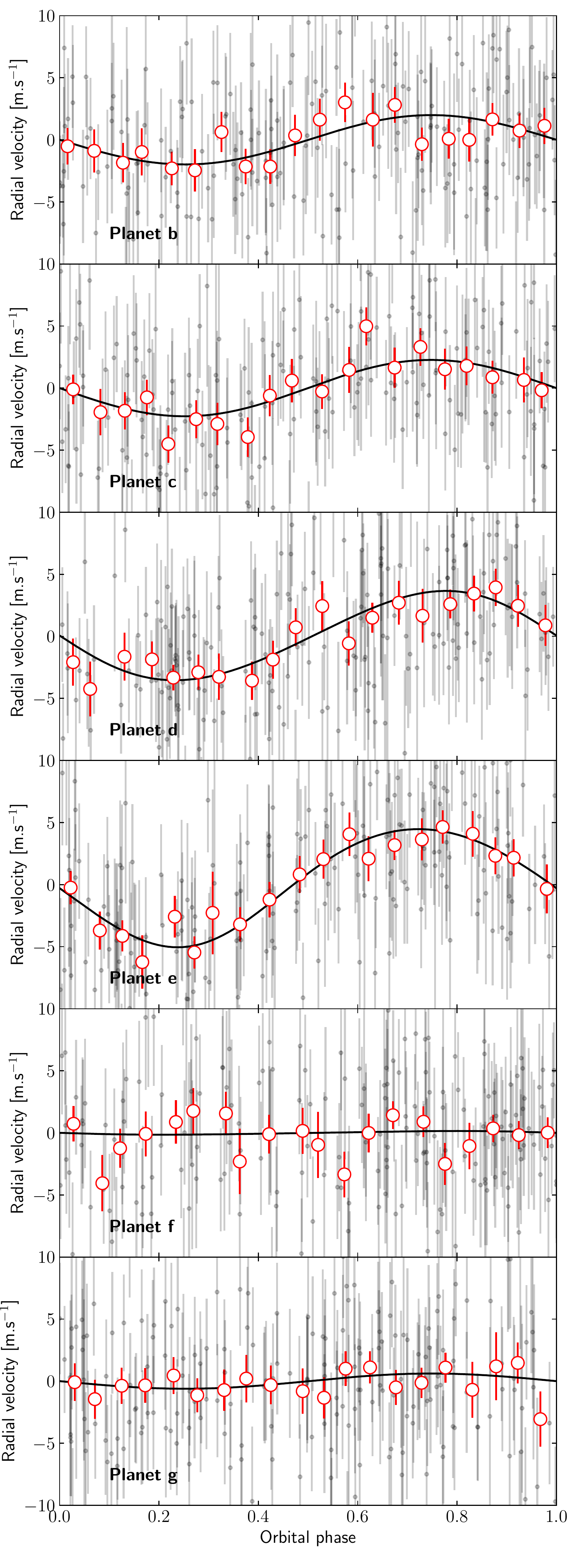} 
	    \caption{\label{rv-phase} HARPS radial velocities in phase at each of the six orbital periods (from top to bottom: planet b to g).}
	\end{figure}

Similarly to previous studies \citep{2017A&A...604A..19O, 2017A&A...608A..25B, 2018NatAs...2..393S, 2018A&A...620A..77L}, we used the bayesian software \texttt{PASTIS} \citep{2014MNRAS.441..983D, 2015MNRAS.451.2337S} to jointly analyse the HARPS radial velocities, the K2 light curve and the spectral energy distribution (SED). For the SED, we used the magnitudes in optical from the APASS survey and in near-infrared from the 2MASS and AllWISE surveys \citep{2015AAS...22533616H, 2014AJ....148...81M, 2014yCat.2328....0C}. The list of magnitudes is given in Table \ref{stellarparam1}. The RVs were modelled with keplerian orbits and the stellar activity with a GP. We did not filtered the activity using de-correlation of the spectroscopic proxies as we did not find any significant correlations between them and the RVs (the highest correlation with RVs is 0.14 for the FWHM). We used a quasi-periodic kernel with the following covariance matrix:

\begin{equation} 
\label{kernelrv}
\mathbf{K}_{\textit{QP}} = A_2^2 ~\exp\left[ -\frac{1}{2} \left( \frac{\mathbf{\Delta t_{\rm RV}}}{\lambda_1} \right) ^2 - \frac{2}{\lambda_2^2} \sin^2 \left( \frac{\pi \left| \mathbf{\Delta t_{\rm RV}} \right| }{P_{rot}} \right) \right] + \mathbf{I} \sqrt{\bm\sigma^2 + \sigma_{j}^2}.
\end{equation} The first term is the quasi-periodic kernel with the hyperparameters: $A_2$ (amplitude), $P_{rot}$ (stellar rotation period), $\lambda_1$ (coherent timescale) and $\lambda_2$ (relative weight between the periodic and decay terms). The matrix $\bold{\Delta t_{\rm RV}}$ have elements $\Delta t_{ij} = t_i - t_j$ from the RV time series. The second term is the identity matrix $\textbf{I}$ multiplied by the data uncertainty vector $\bm\sigma$ plus an extra source of uncorrelated noise $\sigma_j$ (jitter).

The transits were modelled with the \texttt{jktebop} package \citep{2008MNRAS.386.1644S} using an oversampling factor of 30 to account for the long integration time of the data \citep{2010MNRAS.408.1758K}. The SED was modelled using the BT-Settl library of stellar atmosphere models \citep{2012RSPTA.370.2765A}. 

A Markov Chain Monte Carlo (MCMC) method is implemented in \texttt{PASTIS}. It was used to derive the system parameters and their uncertainties. The spectroscopic parameters were converted into physical stellar parameters using the Dartmouth evolution tracks \citep{2008ApJS..178...89D} at each step of the chains. Similarly, the limb darkening coefficients, assuming a quadratic law, were computed using the stellar parameters and tables from \citet{2011A&A...529A..75C}. 

The complete list of priors of fitted parameters is given in Table \ref{MCMCprior} in the Appendix. For the stellar temperature, surface gravity and Iron abundance, we used normal priors centred on the values from the spectral analysis. For the period and transit epoch, we used normal priors centred on the values from \citet{2018AJ....155...57C}. For the systemic distance to Earth, we used a normal prior centred on the \textit{Gaia} Data Release 2 \citep{2018A&A...616A...1G} distance value. For the orbital inclination, we used a sine distribution and for the orbital eccentricity, we used a truncated normal distribution with width $\sigma = 0.083$ as reported by \citet{2019AJ....157...61V} for small multi-transiting exoplanets ($R < 6 \, R_{\oplus}$). For the other parameters, we used uniform uninformative priors with ranges shown in Table \ref{MCMCprior}.

\begin{figure}[htbp!]
	\centering
	    \includegraphics[width=0.47\textwidth]{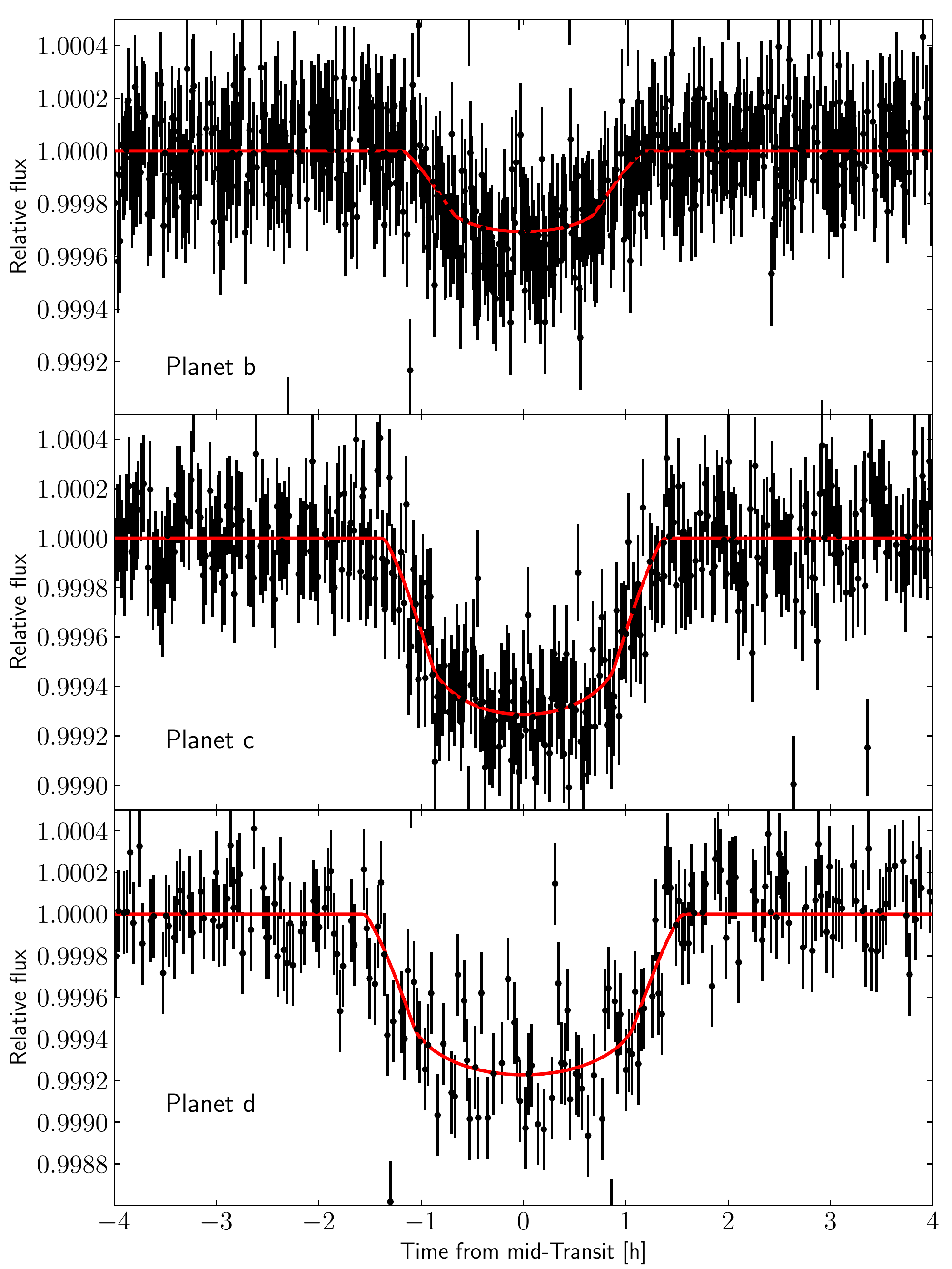} 
	    
        \vspace{-0.05cm}	    
        
	    \includegraphics[width=0.47\textwidth]{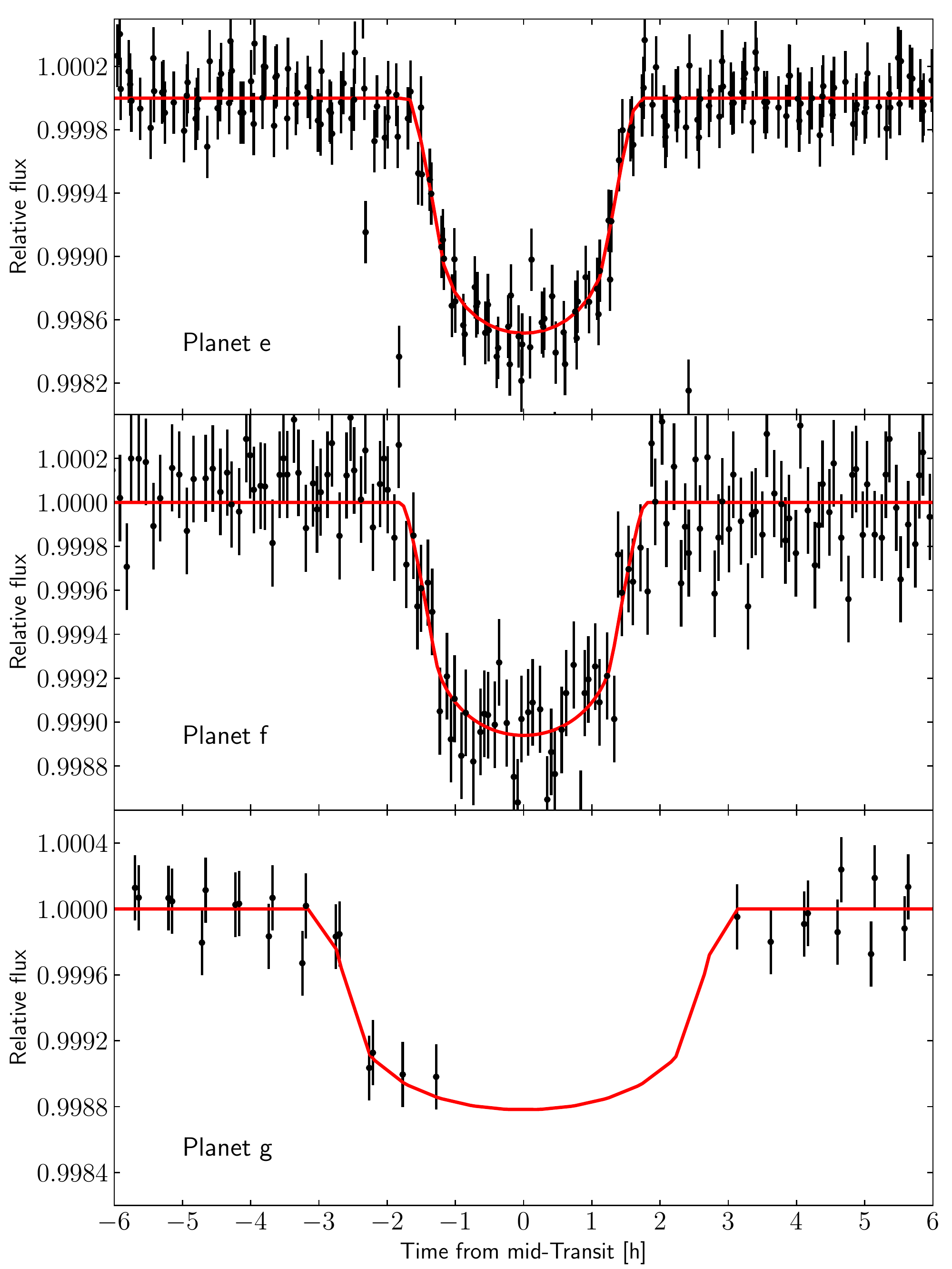} 
	    \caption{\label{transit_in_phase} Phase-folded transit light curves of planets (from top to bottom) b, c, d, e, f, and g. The best transit model is shown in red. The time scale is not the same between the first three and last three plots. The coverage of the last transit is not as good as in \citet{2018AJ....155...57C} due to the differences in the pipeline used to correct for \textit{K2} systematics.}
	\end{figure}

The parameter space was explored in several steps and the priors reported in Table \ref{MCMCprior} are applicable to the last step of the analysis. First, we normalised the light curve for the joint analysis. For this, we ran 20 MCMCs with $3\times 10^5$ iterations on the photometric data alone, modelling the six transiting planets starting from the values reported in \citet{2018AJ....155...57C}, and the stellar modulation with a GP, using the following square exponential kernel:

\begin{equation}
\label{kernelphot}
\mathbf{K}_{\textit{SE}} = A_1^2 ~\exp\left[ -\frac{1}{2} \left( \frac{ \mathbf{\Delta t_{\textit{K2}}}}{l} \right) ^2 \right] + \mathbf{I} \sqrt{\bm \sigma^2 + \sigma_{j}^2}.
\end{equation} The first term is the square exponential kernel with the hyperparameters: $A_1$ (amplitude) and l (coherent time scale). The matrix $\bold{\Delta t_{\textit{K2}}}$ have elements $\Delta t_{ij} = t_i - t_j$ from the photometric time series. The second term is similar to the one in Eq. \ref{kernelrv}. We used a square exponential kernel instead of a quasi-periodic kernel as it requires fewer parameters and the goal at this stage was to fit the light curve and not to derive the stellar rotation period as already done in Sect. \ref{rotation} on the binned light curve. We checked the convergence with a Kolmogorov-Smirnov test, removed the burn-in phase and merged the remaining chains. Then, we subtracted the resulting model of the six planets, removing the transits. Finally, the prediction at observed times on the residuals (activity modulation and residual systematics) of the GP was subtracted from the complete light curve, including the transits. The result is shown in Fig. \ref{k2lc}, bottom part. We used the resulting normalised light curve in the final analysis with radial velocities. For this, we ran 96 MCMCs with $6 \times 10^5$ iterations. We there also checked the convergence with a Kolmogorov-Smirnov test, removed the burn-in phase and merged the remaining chains. We used the results as starting points for a new iteration of the analysis, to increase the number of chains with the same posterior probability distributions. The results of the final run are shown in Table \ref{MCMCprior}, including stellar parameters derived in the joint analysis. The RVs and transits in phase at the periods of each planets are shown respectively in Fig. \ref{rv-phase} and \ref{transit_in_phase}. The radial velocities, together with the six-planet model and the activity GP regression, are shown in Fig. \ref{rv}. We also ran a full analysis using the \texttt{PARSEC} evolution tracks \citep{2012MNRAS.427..127B} to check whether the parameters were consistent, which is the case, as it is also reported in Table \ref{MCMCprior}.

\section{Discussion \label{sect4}}
	
\subsection{Validity of the detections}
\begin{figure*}[h!]	
	    \sidecaption
  \includegraphics[width=12cm]{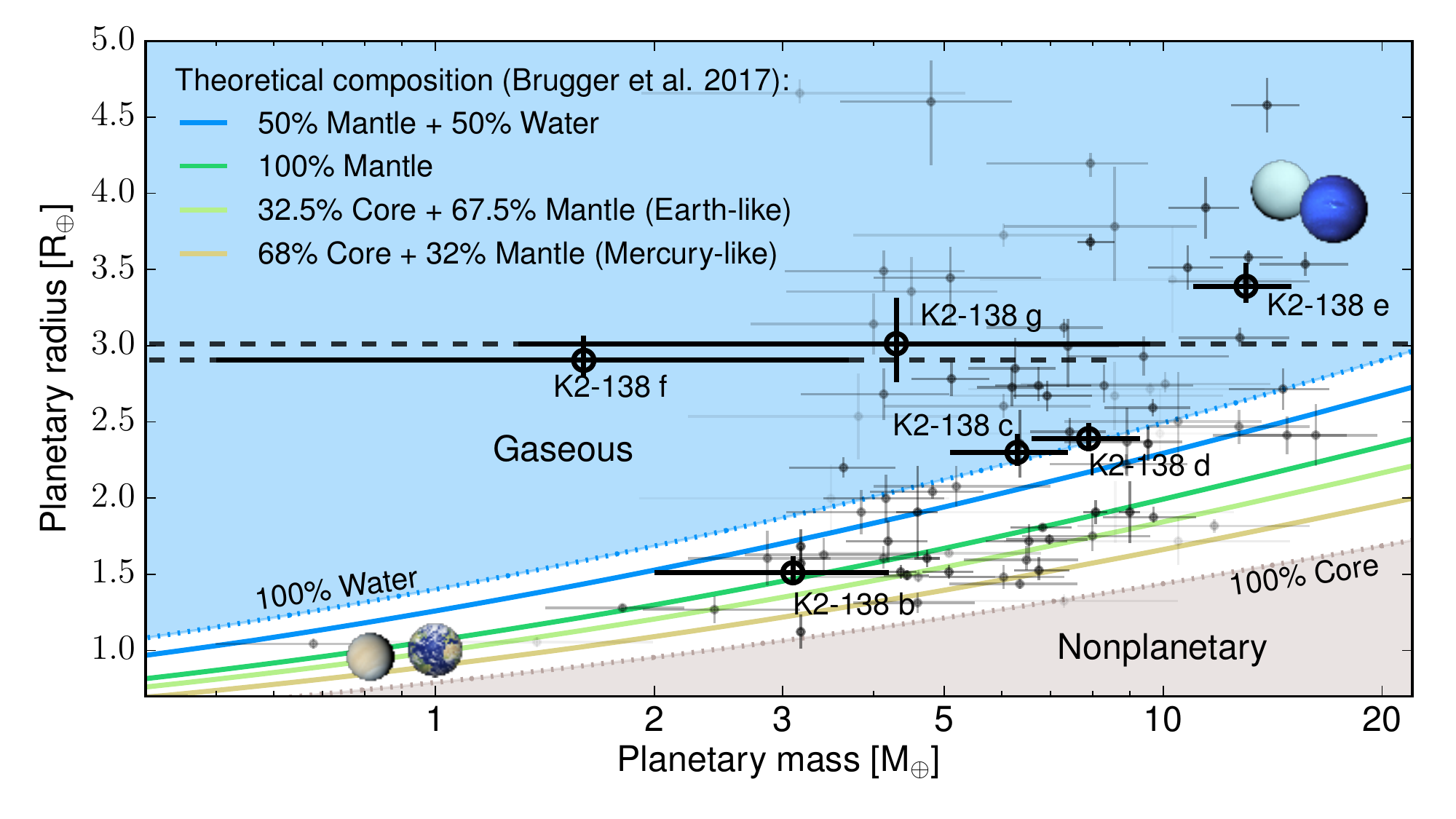}
	    \caption{\label{MRplot} Mass-radius diagram of small planets with masses up to $22\ \rm M_{\oplus}$ and radius up to $5\ \rm R_{\oplus}$. From top to bottom, the lines denotes different compositions for solid planets in between pure water and pure iron \citep{2017ApJ...850...93B}. We superimposed the known planets (from the NASA Exoplanet Archive, updated on 2019 May 05) in this mass-radius range where the grey-scale depends on the precision on the mass and radius. We mark the positions of the planets K2-138b, c, d, e, f and g. The uncertainties are the 68.3\% confidence interval. For planets f and g, which are not detected, we overplot the 99\% confidence interval on the mass in dashed black.}

	\end{figure*}

We were able to derive the masses of planets b, c, d, and e: $\rm M_b =  3.1\pm1.1$ \Mearth, $\rm M_c = 6.3^{_{+1.1}}_{^{-1.2}}$ \Mearth, $\rm M_d =  7.9^{_{+1.4}}_{^{-1.3}}$ \Mearth\, and $\rm M_e =13.0\pm2.0$ \Mearth\ with a precision of 34\%, 20\%, 18\% and 15\%, respectively. The radii are compatible with \citet{2018AJ....155...57C}. The bulk densities are $4.9^{_{+2.0}}_{^{-1.8}}$ \gcm3, $2.8\pm0.7$ \gcm3, $3.2\pm0.7$ \gcm3, and $1.8\pm0.4$ \gcm3, respectively, ranging from Earth to Neptune-like values, as shown in the mass-radius plane in Fig. \ref{MRplot}. This brings constraints on their internal structures. Considering their densities and that their mass are $\lesssim 10$ \Mearth, these planets likely have a rocky core and a substantial atmospheric layer, composed of volatiles. Such a layer is not taken into account in current models of super-Earth interiors \citep{2017ApJ...850...93B}. The modelling of the planets is therefore beyond the scope of this paper and will be the subject of a forthcoming study. The detection of multiple relatively low-density planets around a metallic star is also consistent with previous studies showing an increase in frequency and upper-mass boundary of Neptune-like planets with the metallicity of the host star \citep{2016MNRAS.461.1841C}. For planets f and g, we have upper limits at $99 \%$ of 8.7  \Mearth\ and 25.5 \Mearth\ on the masses, leading to upper limits of 2.1 \gcm3 and 5.1  \gcm3 on the densities.

The activity level in the RVs, as fitted by the GP is $5.6^{_{+2.9}}_{^{-1.5}}$ \ms. Unfortunately, this cannot be compared with the photometric activity level since there is a one-year gap between the photometric and spectroscopic observations and the activity level could have changed. We can exclude a stellar origin for the detected signals as the stellar rotation period does not correspond to the periods of the signals at the exception of planet f which we do not detect as it is likely absorbed by the GP regression, being close to half the rotation period (e.g. \citealt{2018A&A...615A..69D}). As such, the upper-limit provided for planet f is likely to be underestimated, and the mass may be higher. Assuming a density for planet f in the range $[1.79, 4.85]$ \gcm3, which is the density range of the other planets detected in the system, we can estimate masses between $8.0$ \Mearth\ and $21.6$ \Mearth, for the derived radius. This range of values is above the upper-limit we obtained for planet f, in line with a likely absorption of the signal by the GP. In this regard, it could have been beneficial to have simultaneous spectropolarimetric observations to optimise the activity filtering as in \citet{2016MNRAS.461.1465H}. It may have prevented the non-detection of planet f. Besides, we are not able to search for the presence of planets in the gaps of the broken resonance chain for the same reason.

\subsection{Transit timing variations \label{TTV}}
\begin{figure}[h!]
	\centering
	    \includegraphics[width=0.45\textwidth]{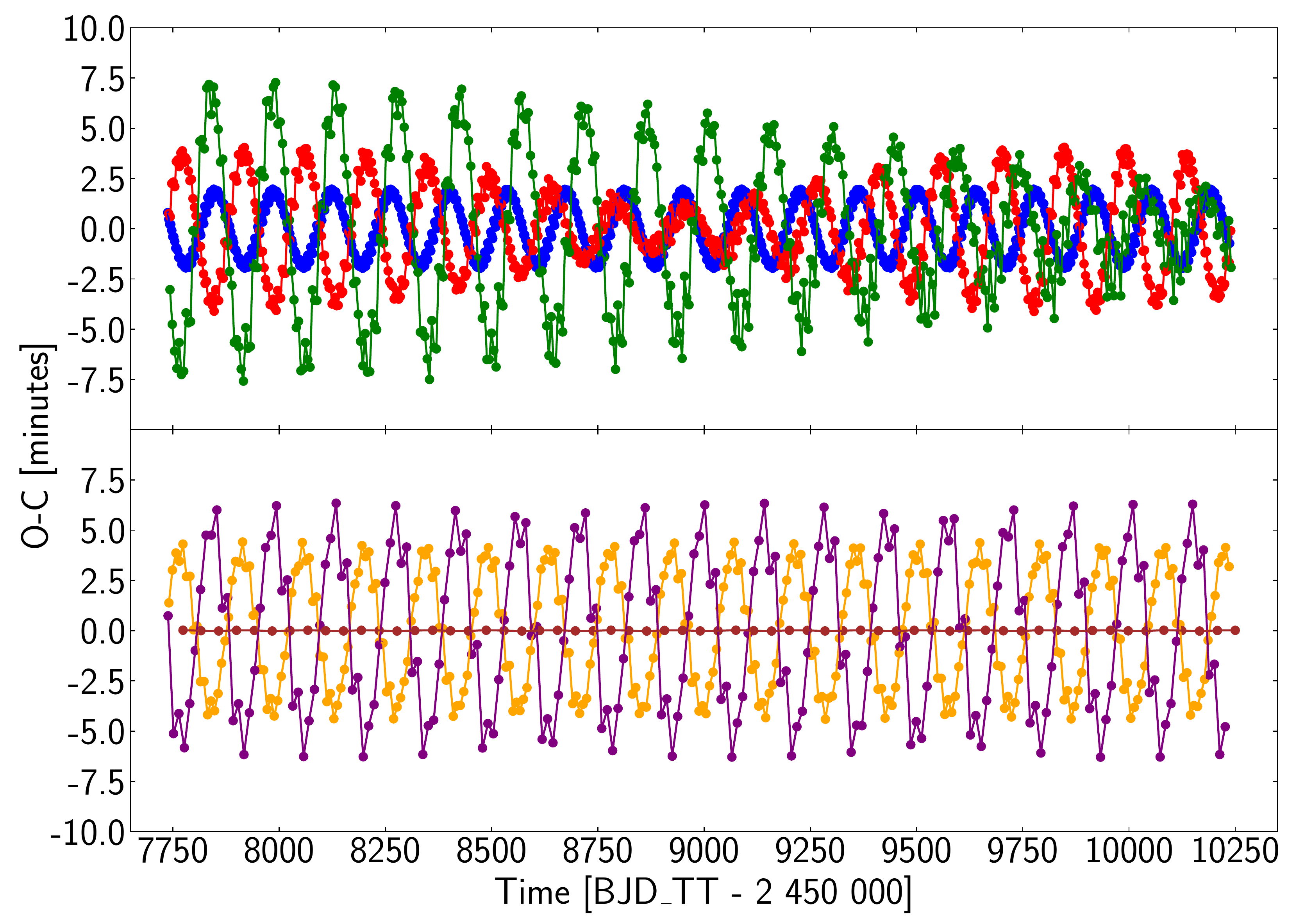}
	    \caption{\label{ttv} Theoretical transit timing variations predicted using \texttt{TTVFaster}, assuming zero eccentricities for all the planets. \textit{Top}: Planet b (in blue), planet c (in red), planet d (in green). \textit{Bottom}: Planet e (in orange), planet f (in purple), planet g (in brown).}	    
	\end{figure}

K2-138 is a remarkable dynamical system with five planets in a chain close to the 3:2 resonance. In addition, these planets form a chain of three-body Laplace resonances \citep{2018AJ....155...57C}. This makes K2-138 an ideal target to study transit timing variations. We estimated the transit timing variations using the \texttt{TTVFaster} code \citep{2016ApJ...818..177A}. We first assumed zero eccentricities for the six planets and took the median masses from Table \ref{K2-138Param}. We found amplitudes of the order of 2.0, 4.1, 7.3, 4.5, 6.4, and 0.02 minutes for planets from b to g, respectively. Then, assuming median eccentricities from Table \ref{K2-138Param}, we obtained amplitudes of the order 34, 42, 66, 37, and 41 minutes for planets b to f (Fig. \ref{ttv}). These amplitudes are excluded by \textit{K2} observations which do not show any significant variations at the level of 8-10 minutes, as demonstrated by \citet{2018AJ....155...57C}. Therefore, the planets are likely close to circular orbits, which is compatible with the posterior distributions we obtained. This is also in line with results showing that tightly packed multi-transiting planets have low eccentricities \citep{2019AJ....157...61V}. Finally, using the upper limits derived for the masses of planets f and g, we computed the corresponding TTVs of planets e and f. Planet g does not impact TTVs of planet f in a significant (detectable) way as it is far from resonance, being located after a double gap in the chain of resonance. Therefore, it is not possible to use TTVs of planet f to constrain the mass of planet g even if there are undetected and non-transiting planets filling the gap in the chain. Still assuming circular orbits, TTVs of planet e reach an amplitude of the order eight minutes only for masses of planet f above 12 \Mearth, which is above the upper limit we derived. However, as we cannot exclude an absorption of the signal of planet f by the GP, it is more conservative to include masses for planet f up to around 12 \Mearth. Using the upper limit mass for planet f, the predicted TTVs of planet e are of the order of 6.4 minutes, and using the median mass they are of 4.5 minutes. Both are well within reach of space mission like CHEOPS which has a cadence of up to 60 seconds. Observations of transits of planet e using CHEOPS would be beneficial to constrain further the mass of planet f through a photodynamical analysis \citep{2015MNRAS.454.4267B}. Additionally, more transit observations, with a higher cadence, of planets b, c and d would allow to constrain the masses of planets c, d and e from TTVs which would make K2-138 an interesting benchmark system for comparing RV and TTV masses. This would allow to better calibrate the two mass measurement techniques. Nevertheless, at this stage, and without more precise photometric observations of planets e and f, an analysis including TTVs would not allow to measure the masses of planets f and g. 

\subsection{A favourable system to search for co-orbitals}
Multi-planetary systems with a relatively large number of planetary components have been studied in theoretical works as potential environments for the formation of co-orbital planet pairs (e.g. \citealt{cresswell06}). In these configurations, two planet-like objects share the same orbital path captured on the gravitational well of each other in different possible 1:1 mean motion resonance (MMR) configurations (e.g. \citealt{laughlin02}). Although theoretical results allow these configurations to exist in nature (e.g. \citealt{laughlin02,cuk12, leleu19}) and different works have searched for them (e.g. \citealt{madhusudhan08, janson13, lillo-box18a, lillo-box18b}), none has yet been detected but some candidates have been announced (e.g.  \citealt{leleu19a}).

The outcome of studies that analyse the dynamical interactions between planetary embryos during the formation and first stages of the evolution of multi-planetary systems suggests that in a relatively large percentage of the systems pairs of embryos end up being captured in these 1:1 resonances. In particular, \cite{cresswell08} found that in 30\% of their simulations the final system contained one pair of co-orbital planets, while \cite{leleu19b} also found this same result for 13\% of their simulated systems (including disk dissipation). The systems found by \cite{leleu19b} ending up in co-orbital configurations were in a vast majority trapped in 3:2 and 4:3 MMR with a third planet and demonstrated that resonant chains can stabilise co-orbital configurations that would be unstable otherwise.

Given its architecture with multiple planets in MMR, K2-138 represents an excellent system to look for these co-orbital configurations in further studies. A dedicated analysis of the \textit{K2} light curve and radial velocity in this regard is, however, out of the scope of this paper and will be studied in future works by following procedures similar to those described in \cite{lillo-box18a} and \cite{janson13}.

\section{Conclusion \label{sect5}}
In this paper, we report the characterisation of the planets around K2-138. Their parameters were derived via a Bayesian combined analysis of both \textit{K2} photometry and HARPS radial-velocities. We computed the following masses for planets c, d and e: $\rm M_c = 6.3^{_{+1.1}}_{^{-1.2}}$ \Mearth, $\rm M_d =  7.9^{_{+1.4}}_{^{-1.3}}$ \Mearth\, and $\rm M_e =13.0\pm2.0$ \Mearth\ with a precision of 20\%, 18\% and 15\%, respectively. For planet b, we have a strong constraint of $\rm M_b =  3.1\pm1.1$ \Mearth\ (precision 34\%). The masses and radii derived lead to bulk densities of $4.9^{_{+2.0}}_{^{-1.8}}$ \gcm3, $2.8\pm0.7$ \gcm3, $3.2\pm0.7$ \gcm3, and $1.8\pm0.4$ \gcm3 ranging from Earth to Neptune-like densities for planet b to e. For the two outer planets, we were not able to detect their masses, mostly due to the level of stellar activity. The upper masses derived are $\rm M_f < 8.7$  \Mearth\ and $\rm M_g < 25.5$ \Mearth\ at 99\% confidence interval, though the one for planet f should be taken with caution as its signal may have been partly absorbed by the activity filtering. The timing precision obtained with \textit{K2} does not allow to use TTVs to bring further constraints on the masses of planets f and g. Additional observations of transits in this system, with a higher cadence than \textit{K2}, would allow both to constrain the mass of planet f and to use K2-138 as a benchmark system to compare RV and TTV masses. Finally, thanks to its dynamical peculiarities, K2-138 is also a good candidate to search for co-orbital bodies.

\begin{acknowledgements}
We thank Jessie L. Christiansen and Kevin Hardegree-Ullman for the useful discussions on TTVs and on the Spitzer transit detection of planet g. We are grateful to the pool of HARPS observers who conducted part of the visitor-mode observations at La Silla Observatory: Julia Seidel, Amaury Triaud, David Martin, Nicola Astudillo, Jorge Martins, Vedad Hodzic. Based on observations made with ESO Telescopes at the La Silla Paranal Observatory under programme ID 198.C-0.168. This publication makes use of The Data \& Analysis Center for Exoplanets (DACE), which is a facility based at the University of Geneva (CH) dedicated to extrasolar planets data visualisation, exchange and analysis. DACE is a platform of the Swiss National Centre of Competence in Research (NCCR) PlanetS, federating the Swiss expertise in Exoplanet research. The DACE platform is available at \url{https://dace.unige.ch}.
This research was made possible through the use of the AAVSO Photometric All-Sky Survey (APASS), funded by the Robert Martin Ayers Sciences Fund.  This publication makes use of data products from the Two Micron All Sky Survey, which is a joint project of the University of Massachusetts and the Infrared Processing and Analysis Center/California Institute of Technology, funded by the National Aeronautics and Space Administration and the National Science Foundation. This publication makes use of data products from the Wide-field Infrared Survey Explorer, which is a joint project of the University of California, Los Angeles, and the Jet Propulsion Laboratory/California Institute of Technology, funded by the National Aeronautics and Space Administration. 
This publication makes use of data products from the Two Micron All Sky Survey, which is a joint project of the University of Massachusetts and the Infrared Processing and Analysis Center California Institute of Technology, funded by the National  Aeronautics and Space Administration and the National Science Foundation.
This paper includes data collected by the K2 mission. Funding for the K2 mission is provided by the NASA Science Mission directorate.
This research has made use of the Exoplanet Follow-up Observation Program website, which is operated by the California Institute of Technology, under contract with the National Aeronautics and Space Administration under the Exoplanet Exploration Program.
This research has made use of NASA's Astrophysics Data System Bibliographic Services.
This work has made use of data from the European Space Agency (ESA) mission
{\it Gaia} (\url{https://www.cosmos.esa.int/gaia}), processed by the {\it Gaia}
Data Processing and Analysis Consortium (DPAC,
\url{https://www.cosmos.esa.int/web/gaia/dpac/consortium}). Funding for the DPAC
has been provided by national institutions, in particular the institutions
participating in the {\it Gaia} Multilateral Agreement.
This research has made use of the VizieR catalogue access tool, CDS, Strasbourg, France. The original description of the VizieR service was published in A\&AS, 143, 23.
DJA acknowledges support from the STFC via an Ernest Rutherford Fellowship (ST/R00384X/1).
SCCB acknowledges support from  Funda\c{c}\~ao para a Ci\^encia e a Tecnologia (FCT) through Investigador FCT contract IF/01312/2014/CP1215/CT0004. O.D.S.D. is supported in the form of work contract (DL 57/2016/CP1364/CT0004) funded by national funds through FCT.
This work was supported by FCT - Fundação para a Ciência e a Tecnologia through national funds and by FEDER - Fundo Europeu de Desenvolvimento Regional through COMPETE2020 - Programa Operacional Competitividade e Internacionalização by these grants: UID/FIS/04434/2019; PTDC/FIS-AST/28953/2017 \& POCI-01-0145-FEDER-028953 and PTDC/FIS-AST/32113/2017 \& POCI-01-0145-FEDER-032113.

\end{acknowledgements}

%
%

\bibliographystyle{aa}
\bibliography{K2-138}

\begin{appendix}
\section{Supplementary Tables and Figures}

\begin{table*}
\centering
\caption{\label{abundance}Chemical abundances of host star, relative to Sun, for main elements with number of lines used for each element.}
\medskip
\resizebox{0.35\textwidth}{!}{
\begin{tabular}{lcc}
\hline
Element & Abundance & Lines number \\
 $[$X/H$]$  & [dex] & \\
\hline
  & &  \\
  {C  \sc   i} &  $  0.18  \pm 0.47$  & 3 \\ 
  {Na \sc   i} &  $  0.17  \pm 0.12$  & 3 \\ 
  {Si \sc   i} &  $  0.16  \pm 0.10$  &  24 \\ 
  {Ca \sc   i} &  $  0.16  \pm 0.10$  & 26  \\ 
  {Ti \sc   i} &  $  0.15  \pm 0.10$   & 43  \\ 
  {Ti \sc  ii} &  $  0.05  \pm 0.11$  & 13  \\ 
  {V  \sc   i} &  $  0.29  \pm 0.10$  & 21 \\ 
  {Cr \sc   i} &  $  0.13  \pm 0.10$  & 19  \\ 
  {Cr \sc  ii} &  $  0.12  \pm 0.12$  & 6  \\ 
  {Fe \sc   i} &  $  0.14  \pm 0.10$  & 235 \\ 
  {Fe \sc  ii} &  $  0.12  \pm 0.10$  & 22  \\ 
  {Co \sc   i} &  $  0.19  \pm 0.10$  & 11  \\ 
  {Ni \sc   i} &  $  0.12  \pm 0.10$  & 67  \\ 
  {Y  \sc  ii} &  $  0.09  \pm 0.11$  & 5  \\ 
\hline
\end{tabular}}
\end{table*}

\begin{figure*}
\centering     
\subfigure[t!][]{\label{rv_dace}\includegraphics[height=0.17\textwidth]{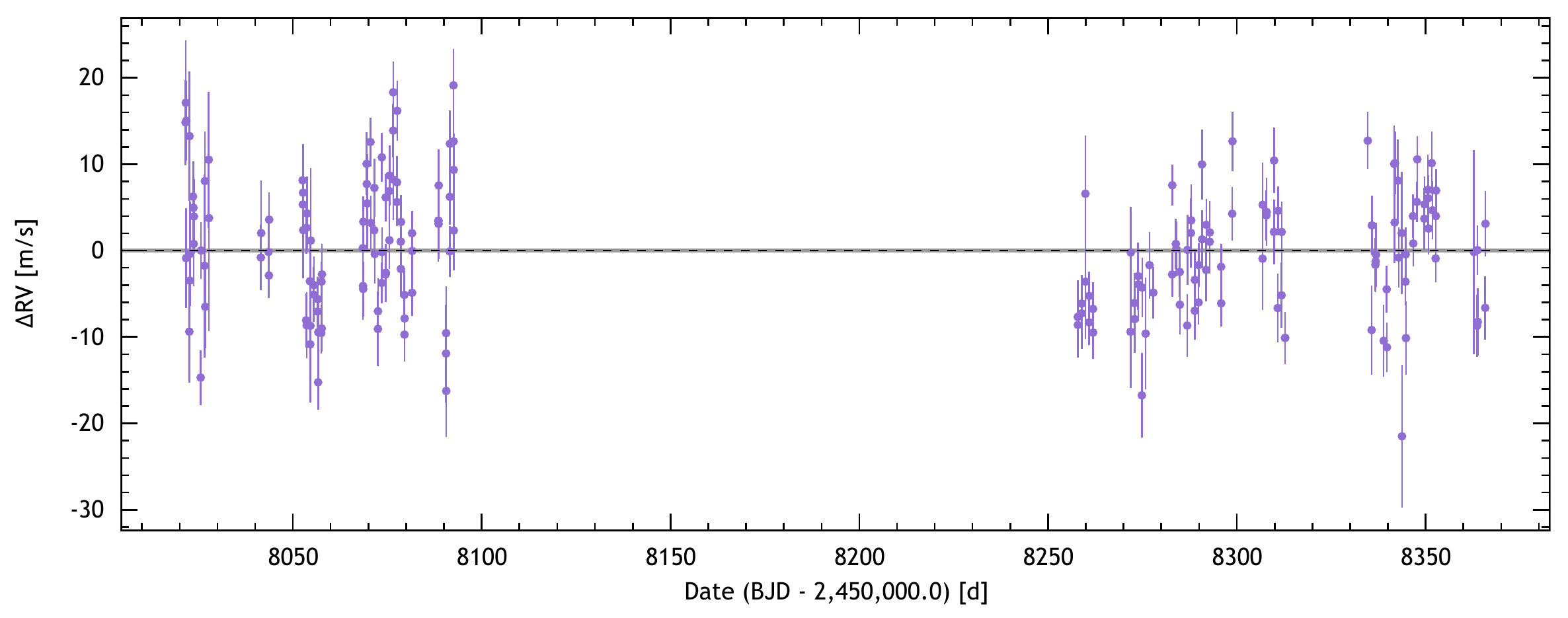}}
\subfigure[t!][]{\label{periodogram_rv_dace}\includegraphics[height=0.17\textwidth]{figures/periodo_rv_dace_1.pdf}}
\subfigure[t!][]{\label{bis_dace}\includegraphics[height=0.17\textwidth]{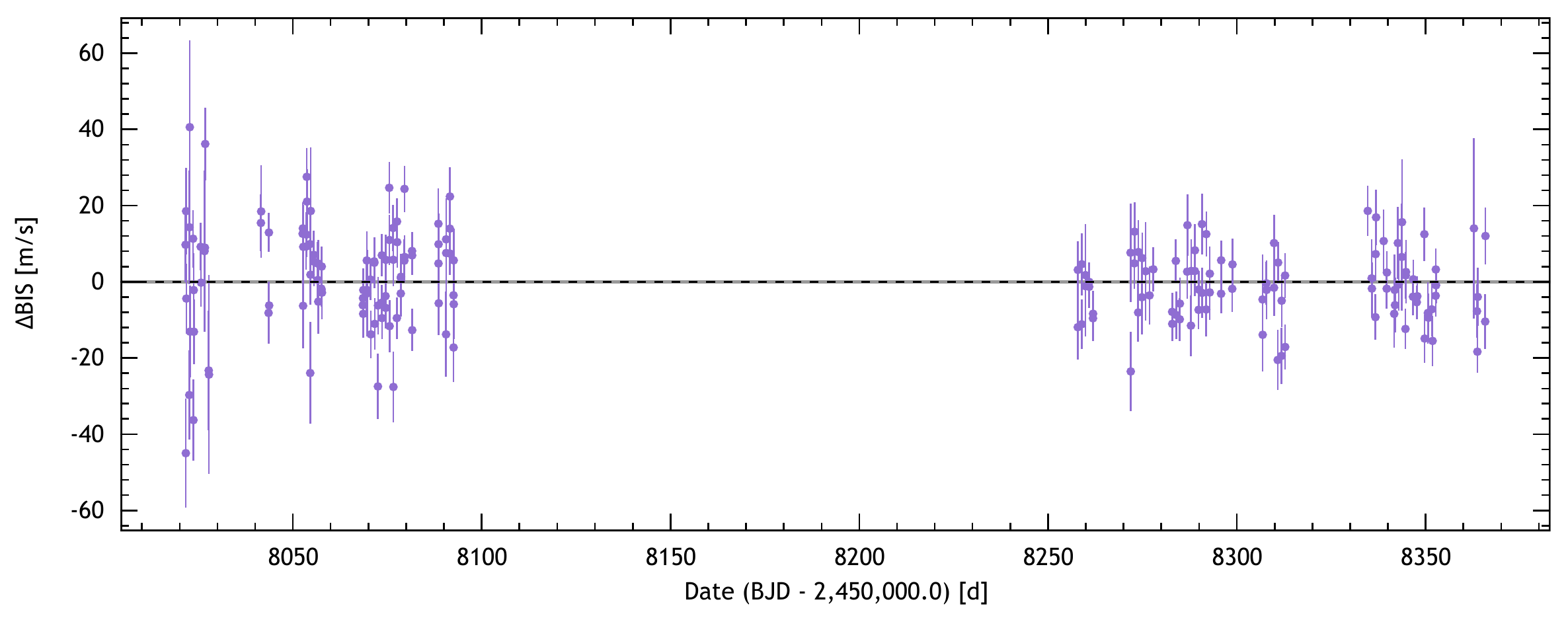}}
\subfigure[t!][]{\label{periodogram_bis_dace}\includegraphics[height=0.17\textwidth]{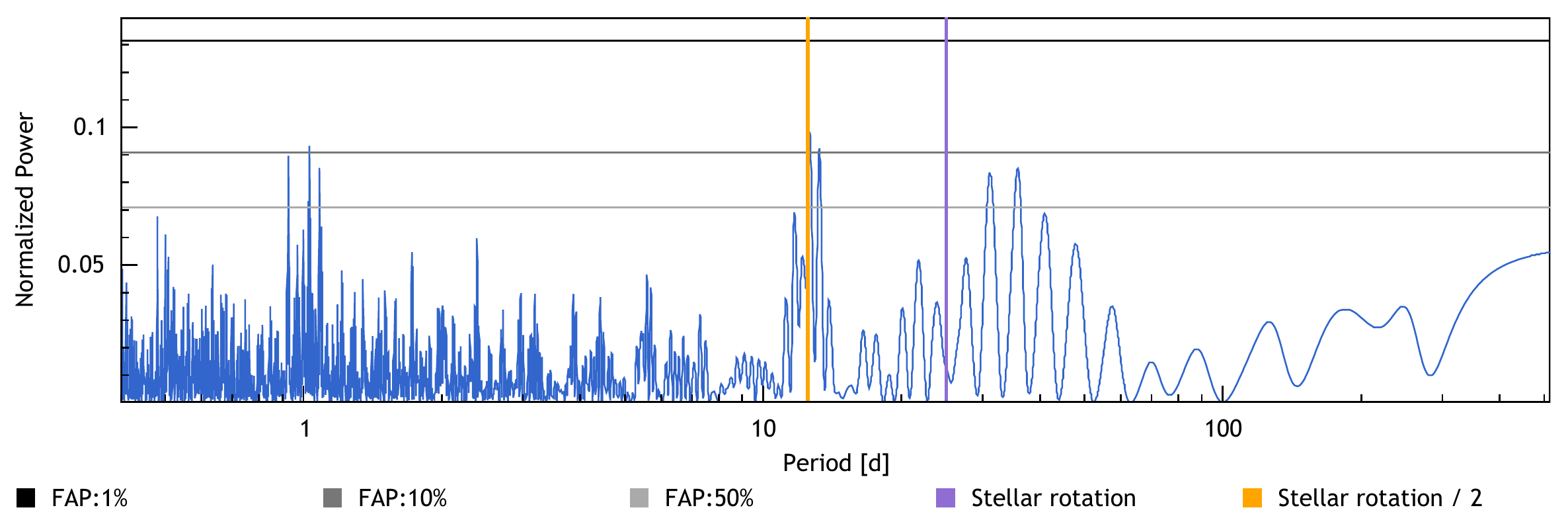}}
\subfigure[t!][]{\label{fwhm_dace}\includegraphics[height=0.17\textwidth]{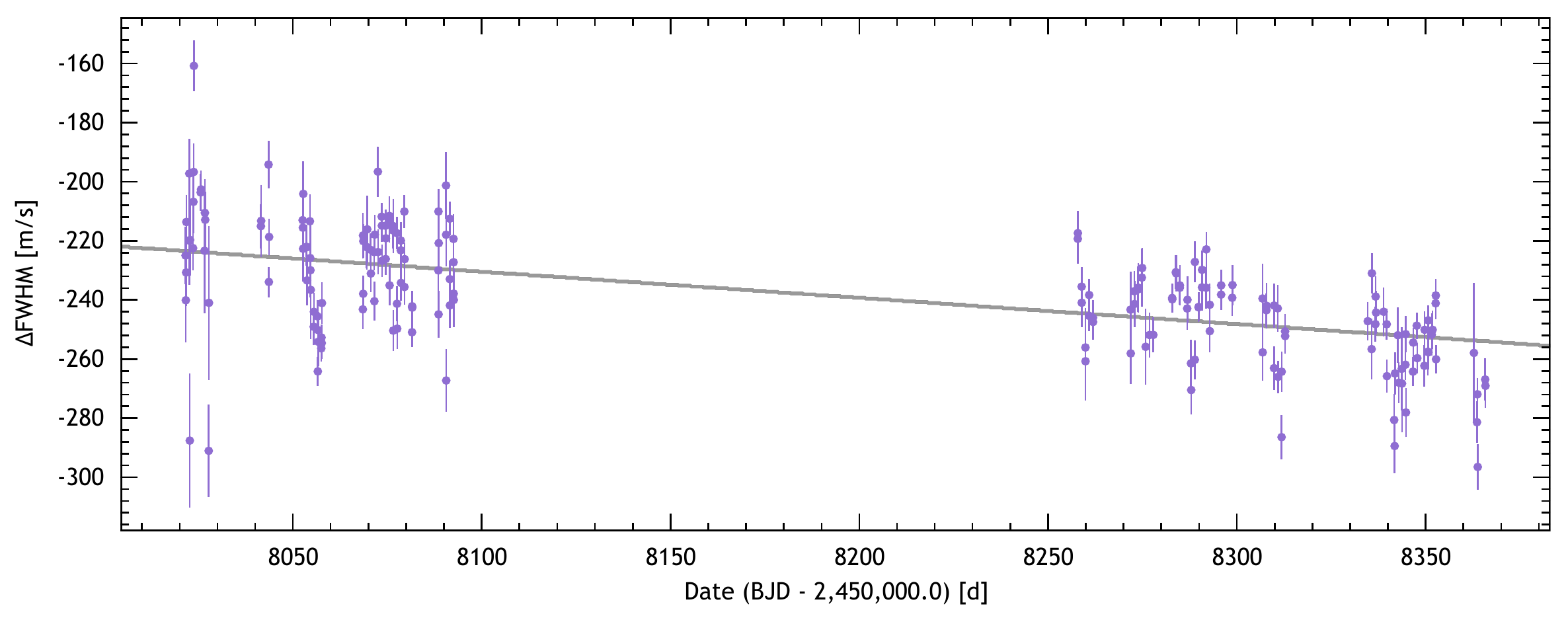}}
\subfigure[t!][]{\label{periodogram_fwhm_dace}\includegraphics[height=0.17\textwidth]{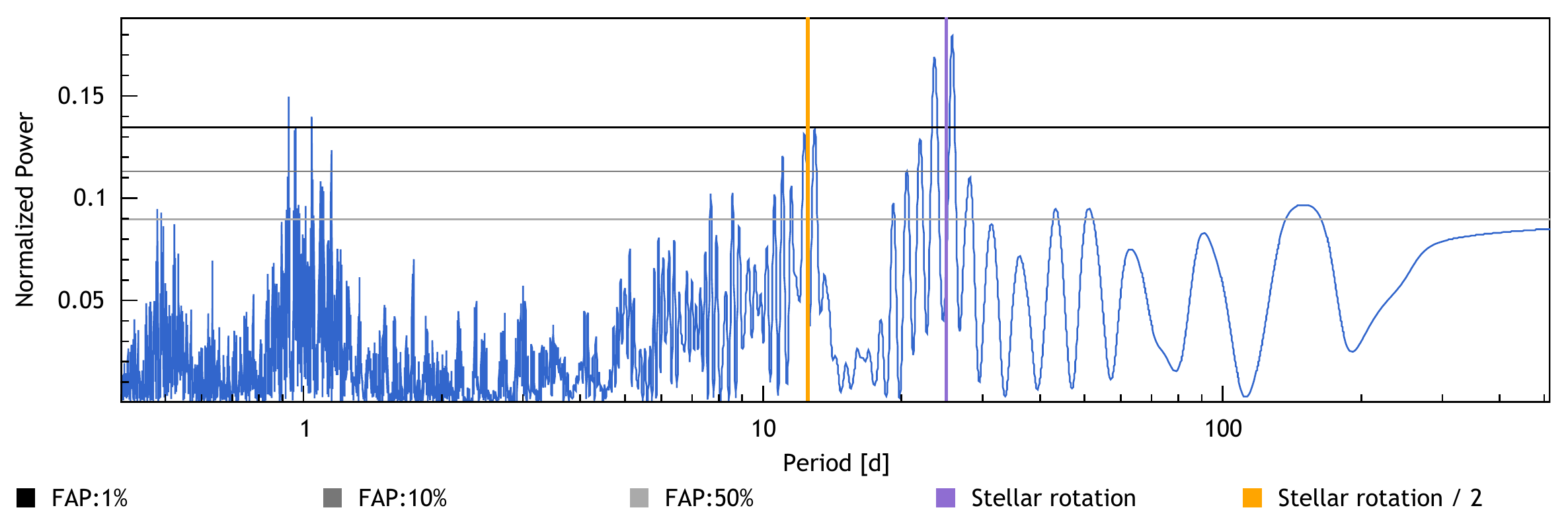}}
\subfigure[t!][]{\label{ha_dace}\includegraphics[height=0.17\textwidth]{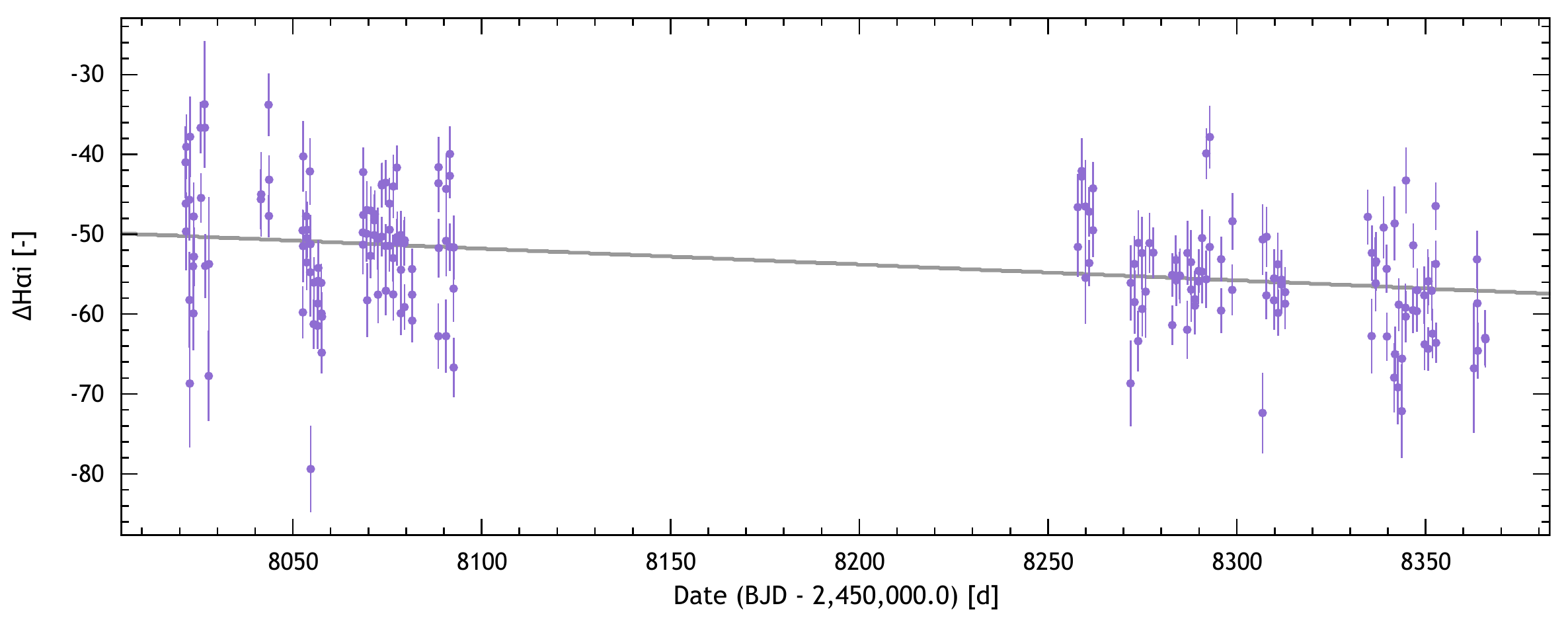}}
\subfigure[t!][]{\label{periodogram_ha_dace}\includegraphics[height=0.17\textwidth]{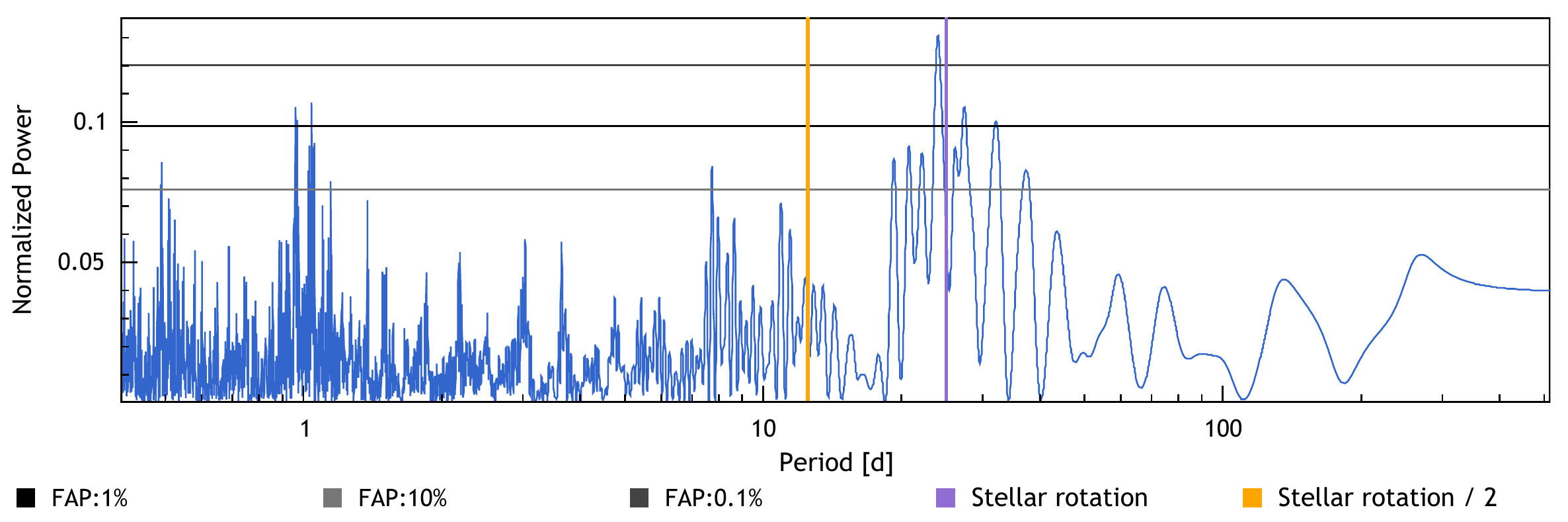}}
\subfigure[t!][]{\label{sindex_dace}\includegraphics[height=0.17\textwidth]{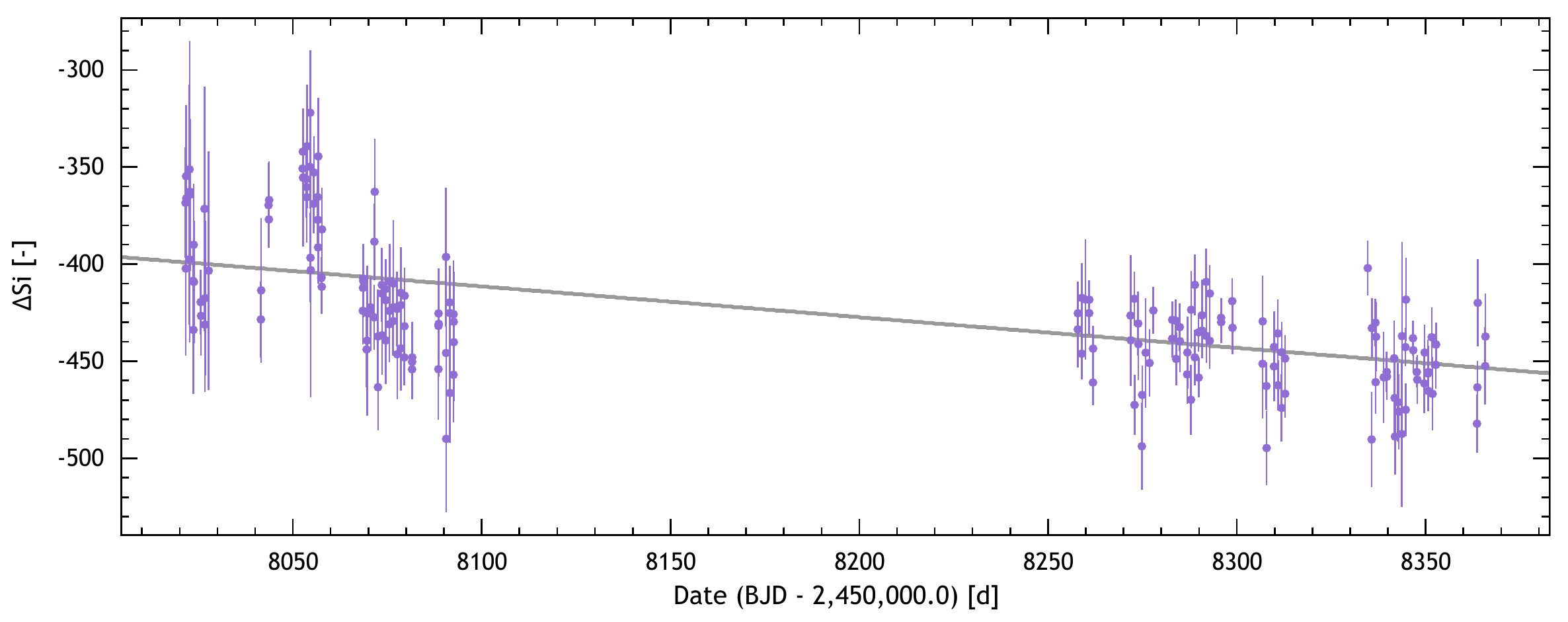}}
\subfigure[t!][]{\label{periodogram_sindex_dace}\includegraphics[height=0.17\textwidth]{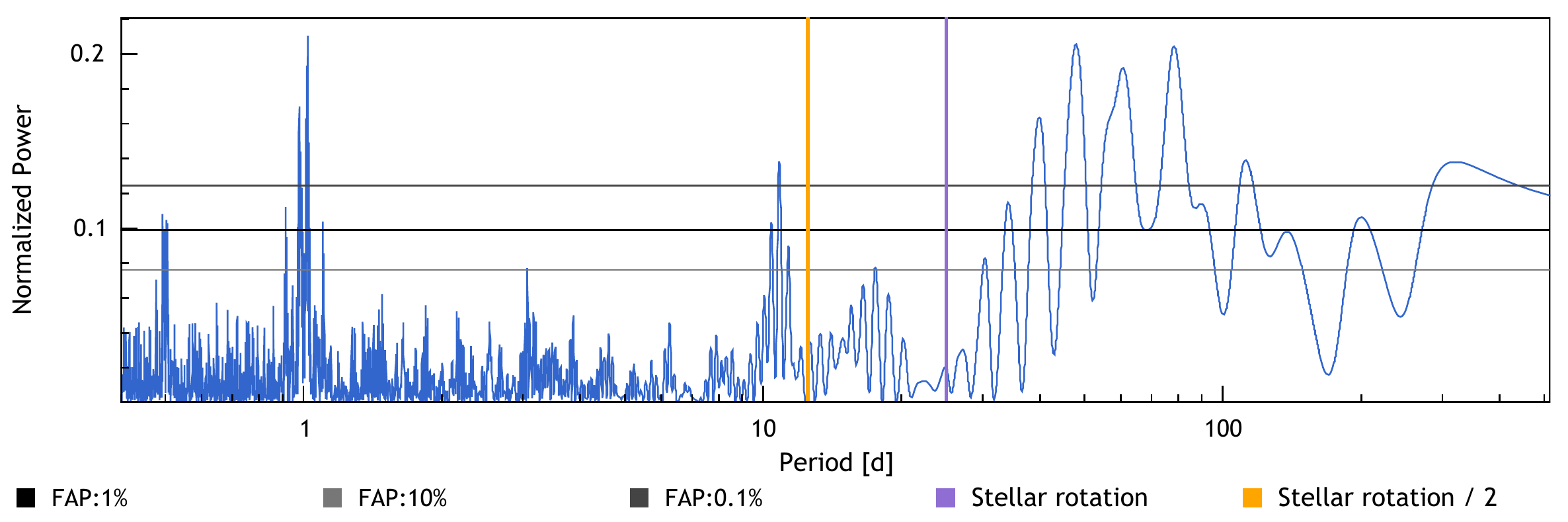}}
\caption{From left to right and top to bottom: Time series and associated Lomb-Scargle periodogram of (a and b) Radial velocity (RV); (c and d) Bisector Span (BIS); (e and f) Full Width Half Maximum (FWHM); (g and h) H$_\alpha$ index; (i and j) S index (S$_{MW}$). The peak position marked by the purple line corresponds to a period of $25$ d (stellar rotation). The orbital periods of the six planets are marked in the RV periodogram. For the FWHM, we fitted a linear drift of $-32.39$ m/s/yr. For the H$_{\alpha}$ index, the drift is $-7.28\pm1.34$. For the S$_{\rm MW}$ index, the drift is $-57.83\pm4.85$ (reference epoch is RJD_TDB 55500 for all of them).}
\label{LSplot}
\end{figure*}

\onecolumn
\begin{landscape}
\footnotesize
\begin{longtable}{lcccccc}
\endfoot
\hline
\\
\multicolumn{7}{l}{}
\endlastfoot
\caption{\label{K2-138Param}System parameters of K2-138 obtained from \texttt{PASTIS}.}\\
\hline 
Parameter &  &  &  &  &  &  \\ 
\hline 
&&&&&& \\
\textit{Stellar parameters} &  &  &  &  &   &  \\ 
&&&&&& \\
Effective temperature \teff\ [K] & $5356^{_{+42}}_{^{-13}}$ &  &  &  &  &  \\ 
Surface gravity \logg\ [cgs] & $4.54^{_{+0.02}}_{^{-0.04}}$ &  &  &  &  &  \\ 
Metallicity [M/H] [dex] & $0.14\pm0.04$  &  &  &  &  &  \\ 
Distance to Earth $D$ [pc] & $208.9^{_{+6.5}}_{^{-4.9}}$  &  &  &  &  &  \\ 
Extinction $E(B-V)$ [mag] & $0.007^{_{+0.01}}_{^{-0.005}}$  &  &  &  &  &  \\ 
Systemic RV $\gamma$ [\kms] & $0.6386^{_{+0.0023}}_{^{-0.0026}}$  &  &  &  &  &  \\ 
Stellar density $\rho_{\star}/\rho_{\astrosun}$ & $1.46^{_{+0.1}}_{^{-0.19}}$  &  &  &  &  &  \\ 
Stellar mass M$_{\star}$\ [\Msun] & $0.93\pm0.02$  &  &  &  &  &  \\ 
Stellar radius R$_{\star}$\ [\Rsun] & $0.86^{_{+0.03}}_{^{-0.02}}$  &  &  &  &  &  \\ 
Stellar age $\tau$\ [Gyr] & $2.8^{_{+3.8}}_{^{-1.7}}$  &  &  &  &  &  \\ 
Limb-darkening $u_{a}$ & $0.496^{_{+0.005}}_{^{-0.010}}$  &  &  &  &  &  \\ 
Limb-darkening $u_{b}$ & $0.205^{_{+0.007}}_{^{-0.003}}$  &  &  &  &  &  \\ 
  &  &  &  &  &  &  \\ 
\hline 
&&&&&& \\
\textit{Planet parameters} & \textit{Planet b} & \textit{Planet c} & \textit{Planet d} & \textit{Planet e} & \textit{Planet f} & \textit{Planet g} \\ 
&&&&&& \\
Orbital Period $P$ [d] & $2.35309\pm0.00022$ & $3.56004^{_{+0.00012}}_{^{-0.00011}}$ & $5.40479\pm0.00021$ & $8.26146^{_{+0.00021}}_{^{-0.00022}}$ & $12.75758^{_{+0.00050}}_{^{-0.00048}}$ & $41.96797^{_{+0.00843}}_{^{-0.00725}}$ \\ 
Transit epoch $T_{0}$ [BJD - 2457700] & $2457773.31683^{_{+0.00091}}_{^{-0.00090}}$ & $2457740.32182^{_{+0.00087}}_{^{-0.00089}}$ & $2457743.15989^{_{+0.00091}}_{^{-0.00093}}$ & $2457740.64558^{_{+0.00085}}_{^{-0.00081}}$ & $2457738.70235^{_{+0.00093}}_{^{-0.00092}}$ & $2457773.86156^{_{+0.01863}}_{^{-0.03219}}$ \\ 
RV semi-amplitude $K$ [\kms] & $0.00156\pm0.00053$ & $0.00278^{_{+0.00050}}_{^{-0.00055}}$ & $0.00303\pm0.00052$ & $0.00431\pm0.00067$ & $0.00047^{_{+0.00061}}_{^{-0.00034}}$ & $0.00083^{_{+0.00102}}_{^{-0.00058}}$ \\ 
 & $[0.000153, 0.003060]$ & $[0.001349, 0.004154]$ & $[0.001585, 0.004314]$ & $[0.002296, 0.006098]$ & $<0.002392$ & $<0.004539$ \\ 
Orbital inclination $i$ [$^{\circ}$] & $87.2^{_{+1.2}}_{^{-1.0}}$ & $88.1\pm0.7$ & $89.0\pm0.6$ & $88.6\pm0.3$ & $88.8\pm0.2$ & $89.4^{_{+0.4}}_{^{-0.3}}$ \\ 
Planet-to-star radius ratio $k$ & $0.01600^{_{+0.00083}}_{^{-0.00076}}$ & $0.02439^{_{+0.00065}}_{^{-0.00059}}$ & $0.02532^{_{+0.00067}}_{^{-0.00063}}$ & $0.03599^{_{+0.00081}}_{^{-0.00073}}$ & $0.03085^{_{+0.00088}}_{^{-0.00089}}$ & $0.03187^{_{+0.00295}}_{^{-0.00250}}$ \\ 
Orbital eccentricity $e$ & $0.048^{_{+0.054}}_{^{-0.033}}$ & $0.045^{_{+0.051}}_{^{-0.032}}$ & $0.043^{_{+0.041}}_{^{-0.030}}$ & $0.077^{_{+0.048}}_{^{-0.049}}$ & $0.062^{_{+0.064}}_{^{-0.043}}$ & $0.059^{_{+0.063}}_{^{-0.040}}$ \\ 
 & $<0.1887$ & $<0.1917$ & $<0.1441$ & $<0.1829$ & $<0.2098$ & $<0.2256$ \\ 
Argument of periastron $\omega$ [$^{\circ}$] & $244^{_{+32}}_{^{-97}}$ & $200^{_{+60}}_{^{-27}}$ & $83^{_{+82}}_{^{-38}}$ & $180^{_{+40}}_{^{-44}}$ & $144^{_{+117}}_{^{-80}}$ & $164^{_{+92}}_{^{-72}}$ \\ 
System scale $a_{b}/R_{\star}$ & $8.46^{_{+0.18}}_{^{-0.37}}$ & $11.14^{_{+0.24}}_{^{-0.49}}$ & $14.72^{_{+0.32}}_{^{-0.65}}$ & $19.53^{_{+0.42}}_{^{-0.86}}$ & $26.10^{_{+0.56}}_{^{-1.15}}$ & $57.73^{_{+1.24}}_{^{-2.55}}$ \\ 
Impact parameter $b$ & $0.42^{_{+0.16}}_{^{-0.18}}$ & $0.37^{_{+0.13}}_{^{-0.14}}$ & $0.24^{_{+0.16}}_{^{-0.15}}$ & $0.47^{_{+0.10}}_{^{-0.13}}$ & $0.56^{_{+0.08}}_{^{-0.11}}$ & $0.56^{_{+0.28}}_{^{-0.36}}$ \\ 
Transit duration T$_{14}$ [h] & $2.010^{_{+0.116}}_{^{-0.122}}$ & $2.385^{_{+0.051}}_{^{-0.060}}$ & $2.703^{_{+0.073}}_{^{-0.072}}$ & $2.974^{_{+0.051}}_{^{-0.052}}$ & $3.205^{_{+0.082}}_{^{-0.077}}$ & $4.751^{_{+0.801}}_{^{-1.415}}$ \\ 
Semi-major axis $a$ [AU] & $0.03385^{_{+0.00023}}_{^{-0.00029}}$ & $0.04461^{_{+0.00030}}_{^{-0.00038}}$ & $0.05893^{_{+0.00040}}_{^{-0.00050}}$ & $0.07820^{_{+0.00053}}_{^{-0.00066}}$ & $0.10447^{_{+0.00070}}_{^{-0.00088}}$ & $0.23109^{_{+0.00154}}_{^{-0.00196}}$ \\ 
Planet mass M [\Mearth] & $3.1\pm1.1$ & $6.3^{_{+1.1}}_{^{-1.2}}$ & $7.9^{_{+1.4}}_{^{-1.3}}$ & $13.0\pm2.0$ & $1.6^{_{+2.1}}_{^{-1.2}}$ & $4.3^{_{+5.3}}_{^{-3.0}}$ \\ 
 & $[0.29, 5.84]$ & $[3.24, 9.44]$ & $[4.14, 11.29]$ & $[7.86, 18.76]$ & $<8.69$ & $<25.47$ \\ 
Planet radius R [\Rearth] & $1.510^{_{+0.110}}_{^{-0.084}}$ & $2.299^{_{+0.120}}_{^{-0.087}}$ & $2.390^{_{+0.104}}_{^{-0.084}}$ & $3.390^{_{+0.156}}_{^{-0.110}}$ & $2.904^{_{+0.164}}_{^{-0.111}}$ & $3.013^{_{+0.303}}_{^{-0.251}}$ \\ 
Planet bulk density $\rho$ [\gcm3] & $4.85^{_{+1.98}}_{^{-1.75}}$ & $2.79^{_{+0.67}}_{^{-0.61}}$ & $3.15^{_{+0.69}}_{^{-0.60}}$ & $1.79^{_{+0.36}}_{^{-0.32}}$ & $0.35^{_{+0.49}}_{^{-0.26}}$ & $0.86^{_{+1.07}}_{^{-0.60}}$ \\ 
 & $[0.30, 10.61]$ & $[1.347, 4.611]$ & $[1.605, 4.960]$ & $[0.986, 2.930]$ & $<2.068$ & $<5.056$ \\ 
Mean equilibrium temperature $T_{eq}$ [K] & $1308^{_{+24}}_{^{-20}}$ & $1140^{_{+21}}_{^{-17}}$ & $992^{_{+18}}_{^{-15}}$ & $861^{_{+16}}_{^{-13}}$ & $745^{_{+14}}_{^{-11}}$ & $501^{_{+9}}_{^{-7}}$ \\ 
Day-side temperature $T_{eq, d}$ [K] & $1346^{_{+41}}_{^{-34}}$ & $1169^{_{+38}}_{^{-27}}$ & $1016^{_{+32}}_{^{-24}}$ & $898^{_{+29}}_{^{-28}}$ & $771^{_{+31}}_{^{-21}}$ & $518^{_{+20}}_{^{-13}}$ \\ 
&&&&&& \\
\hline 
\end{longtable}
\tablefoot{Stellar parameters given here were derived as described in section \ref{companion_star} and are not from the spectral analysis.  For the RV semi-amplitude, orbital eccentricity, planet mass and planet bulk density, the 99\% credible intervals are given on the second lines. We assumed $\Rsun = 695\:508$ km, $\Msun = 1.98842 \times 10^{30}$ kg, $\Rearth = 6\:378\:137$ m, $\Mearth = 5.9736 \times 10^{24}$ kg and 1 AU $= 149\:597\:870.7$ km. The temperatures were derived assuming a zero albedo. The day-side temperature was computed assuming tidally synchronised rotation.}
\end{landscape}
\normalsize

\newpage
\onecolumn

\begin{longtable}{ccccccccccc}
\caption{\label{RVtable}\textbf{Radial velocity data.} The Barycentric Julian Date (BJD) is given with an offset of 2400000. Signal-to-noise ratio (S/N) is given per CCD pixel at 550 nm.}\\
\hline
Time & RV & $\sigma$RV & FWHM & $\sigma$FWHM & BIS & $\sigma$BIS & S$_{\rm MW}$ & $\sigma$S$_{\rm MW}$ & Texp & S/N\\
$[$BJD_UTC$]$ & [\kms] & [\ms] & [\kms] & [\ms] & [\ms] & [\ms]  & & & [s] & \\
\hline
\endfirsthead
\multicolumn{10}{l}{{\bfseries \tablename\ \thetable{} -- continued from previous page}} \\
\hline
Time & RV & $\sigma$RV & FWHM & $\sigma$FWHM & BIS & $\sigma$BIS & S$_{\rm MW}$ & $\sigma$S$_{\rm MW}$ & Texp & S/N\\
\hline
\endhead
\multicolumn{11}{l}{{Continued on next page}} \\ 
\hline
\endfoot
\hline
\endlastfoot
58021.57307 & 0.65402 & 4.89 & 7.0944 & 9.8 & 2.5 & 9.8 & 0.3246 & 0.0284 & 1800 & 22.7 \\
58021.63898 & 0.65629 & 7.16 & 7.0792 & 14.3 & -52.2 & 14.3 & 0.2908 & 0.0449 & 1800 & 17.1 \\
58021.71685 & 0.63831 & 5.70 & 7.0886 & 11.4 & 11.4 & 11.4 & 0.3384 & 0.0365 & 1800 & 20.3 \\
58021.77973 & 0.65421 & 4.59 & 7.1057 & 9.2 & -11.6 & 9.2 & 0.3270 & 0.0323 & 1800 & 24.4 \\
58022.55618 & 0.62980 & 5.87 & 7.1221 & 11.7 & -36.9 & 11.7 & 0.2956 & 0.0335 & 1800 & 19.7 \\
58022.62187 & 0.65243 & 7.45 & 7.0994 & 14.9 & 7.1 & 14.9 & 0.3419 & 0.0437 & 1800 & 16.3 \\
58022.68938 & 0.63570 & 11.36 & 7.0318 & 22.7 & 33.4 & 22.7 & 0.3304 & 0.0777 & 1800 & 11.6 \\
58022.76742 & 0.63880 & 6.00 & 7.0997 & 12.0 & -20.3 & 12.0 & 0.3288 & 0.0391 & 1800 & 19.8 \\
58023.55604 & 0.64543 & 3.75 & 7.0969 & 7.5 & 4.1 & 7.5 & 0.2845 & 0.0197 & 1800 & 27.5 \\
58023.65227 & 0.64414 & 5.29 & 7.1125 & 10.6 & -43.5 & 10.6 & 0.2592 & 0.0329 & 1800 & 21.3 \\
58023.72421 & 0.63995 & 4.86 & 7.1227 & 9.7 & -9.4 & 9.7 & 0.3031 & 0.0315 & 1800 & 22.9 \\
58023.77821 & 0.64313 & 4.26 & 7.1585 & 8.5 & -20.3 & 8.5 & 0.2839 & 0.0316 & 1800 & 25.8 \\
58025.57317 & 0.62448 & 3.09 & 7.1157 & 6.2 & 2.0 & 6.2 & 0.2735 & 0.0167 & 1800 & 32.3 \\
58025.67110 & 0.63919 & 3.18 & 7.1168 & 6.4 & -7.4 & 6.4 & 0.2665 & 0.0203 & 1800 & 32.3 \\
58026.59379 & 0.63742 & 10.57 & 7.0960 & 21.1 & 0.9 & 21.1 & 0.3216 & 0.0629 & 1800 & 12.5 \\
58026.69025 & 0.64723 & 5.69 & 7.1088 & 11.4 & 1.8 & 11.4 & 0.2620 & 0.0347 & 1800 & 20.3 \\
58026.77234 & 0.63268 & 4.78 & 7.1065 & 9.6 & 29.0 & 9.6 & 0.2756 & 0.0397 & 1800 & 23.8 \\
58027.70118 & 0.64970 & 7.83 & 7.0283 & 15.7 & -30.5 & 15.7 & 0.2897 & 0.0613 & 1800 & 16.1 \\
58027.77110 & 0.64294 & 13.03 & 7.0784 & 26.1 & -31.5 & 26.1 & 0.1212 & 0.0955 & 1800 & 11.2 \\
58041.52543 & 0.63838 & 3.72 & 7.1043 & 7.4 & 8.3 & 7.4 & 0.2646 & 0.0196 & 1800 & 27.7 \\
58041.59452 & 0.64121 & 6.08 & 7.1061 & 12.2 & 11.3 & 12.2 & 0.2796 & 0.0372 & 1156 & 19.1 \\
58043.54768 & 0.63902 & 4.02 & 7.1251 & 8.0 & -15.4 & 8.0 & 0.3235 & 0.0218 & 1800 & 26.4 \\
58043.64113 & 0.63630 & 2.54 & 7.0854 & 5.1 & 5.7 & 5.1 & 0.3161 & 0.0143 & 1800 & 38.8 \\
58043.70259 & 0.64276 & 3.11 & 7.1007 & 6.2 & -13.4 & 6.2 & 0.3261 & 0.0199 & 1800 & 32.8 \\
58052.57655 & 0.64732 & 2.51 & 7.1063 & 5.0 & 5.4 & 5.0 & 0.3423 & 0.0130 & 1800 & 38.5 \\
58052.63828 & 0.64452 & 3.43 & 7.1038 & 6.9 & 6.8 & 6.9 & 0.3510 & 0.0198 & 1800 & 29.9 \\
58052.69670 & 0.64588 & 5.59 & 7.0967 & 11.2 & -13.5 & 11.2 & 0.3376 & 0.0348 & 1800 & 20.7 \\
58052.74297 & 0.64155 & 5.49 & 7.1152 & 11.0 & 2.0 & 11.0 & 0.3378 & 0.0356 & 1800 & 21.1 \\
58053.50971 & 0.63109 & 3.04 & 7.0971 & 6.1 & 2.1 & 6.1 & 0.3369 & 0.0146 & 1800 & 32.6 \\
58053.59284 & 0.64182 & 2.92 & 7.0973 & 5.8 & 5.2 & 5.8 & 0.3328 & 0.0158 & 1800 & 34.0 \\
58053.66782 & 0.63053 & 3.78 & 7.0972 & 7.6 & 20.4 & 7.6 & 0.3276 & 0.0235 & 1800 & 27.7 \\
58053.74614 & 0.64348 & 4.19 & 7.0860 & 8.4 & 13.9 & 8.4 & 0.3538 & 0.0318 & 1800 & 25.9 \\
58054.50380 & 0.63563 & 4.52 & 7.1060 & 9.0 & 2.7 & 9.0 & 0.3432 & 0.0217 & 1800 & 23.8 \\
58054.58481 & 0.62833 & 6.67 & 7.0935 & 13.3 & -31.1 & 13.3 & 0.3711 & 0.0321 & 1800 & 17.5 \\
58054.65063 & 0.63048 & 3.92 & 7.0894 & 7.8 & -5.3 & 7.8 & 0.2964 & 0.0229 & 1800 & 26.8 \\
58054.72454 & 0.64034 & 8.35 & 7.0827 & 16.7 & 11.4 & 16.7 & 0.2900 & 0.0653 & 1800 & 15.5 \\
58055.50746 & 0.63408 & 3.10 & 7.0703 & 6.2 & -0.1 & 6.2 & 0.3242 & 0.0151 & 1800 & 31.9 \\
58055.56092 & 0.63519 & 3.24 & 7.0753 & 6.5 & -1.9 & 6.5 & 0.3403 & 0.0187 & 1200 & 31.3 \\
58056.51406 & 0.63213 & 2.78 & 7.0738 & 5.6 & -2.4 & 5.6 & 0.3277 & 0.0136 & 1800 & 35.1 \\
58056.59736 & 0.63356 & 2.46 & 7.0552 & 4.9 & -6.7 & 4.9 & 0.3159 & 0.0145 & 1800 & 39.4 \\
58056.67119 & 0.62393 & 3.09 & 7.0692 & 6.2 & -2.4 & 6.2 & 0.3017 & 0.0185 & 1800 & 32.6 \\
58056.72247 & 0.62973 & 4.17 & 7.0652 & 8.3 & -12.4 & 8.3 & 0.3486 & 0.0301 & 1800 & 26.2 \\
58057.51312 & 0.62964 & 2.23 & 7.0630 & 4.5 & -9.0 & 4.5 & 0.2859 & 0.0098 & 1800 & 42.8 \\
58057.57375 & 0.63559 & 2.22 & 7.0666 & 4.4 & -9.2 & 4.4 & 0.2863 & 0.0103 & 1800 & 43.5 \\
58057.63339 & 0.63017 & 2.61 & 7.0648 & 5.2 & -3.2 & 5.2 & 0.2814 & 0.0140 & 1800 & 38.4 \\
58057.69611 & 0.63642 & 3.48 & 7.0783 & 7.0 & -10.0 & 7.0 & 0.3110 & 0.0215 & 1500 & 30.8 \\
58068.51855 & 0.63946 & 3.31 & 7.0762 & 6.6 & -13.3 & 6.6 & 0.2691 & 0.0172 & 1800 & 30.0 \\
58068.56723 & 0.63508 & 3.83 & 7.1012 & 7.7 & -11.5 & 7.7 & 0.2809 & 0.0211 & 1800 & 26.6 \\
58068.61196 & 0.63473 & 3.08 & 7.0814 & 6.2 & -15.6 & 6.2 & 0.2858 & 0.0179 & 1800 & 32.2 \\
58068.65218 & 0.64252 & 2.84 & 7.0992 & 5.7 & -9.4 & 5.7 & 0.2844 & 0.0185 & 1800 & 35.1 \\
58069.52637 & 0.64922 & 3.55 & 7.0972 & 7.1 & -9.3 & 7.1 & 0.2492 & 0.0195 & 1800 & 28.5 \\
58069.57354 & 0.64689 & 3.78 & 7.0972 & 7.6 & -1.6 & 7.6 & 0.2680 & 0.0214 & 1800 & 27.0 \\
58069.68297 & 0.64464 & 5.66 & 7.1033 & 11.3 & -10.2 & 11.3 & 0.2537 & 0.0387 & 1800 & 20.2 \\
58070.53804 & 0.65175 & 2.71 & 7.0963 & 5.4 & -6.6 & 5.4 & 0.2709 & 0.0150 & 1800 & 35.8 \\
58070.64141 & 0.64239 & 3.09 & 7.0883 & 6.2 & -21.0 & 6.2 & 0.2675 & 0.0190 & 1800 & 32.4 \\
58071.52559 & 0.64153 & 3.03 & 7.0956 & 6.1 & -1.8 & 6.1 & 0.2657 & 0.0167 & 1800 & 32.4 \\
58071.57297 & 0.64643 & 3.29 & 7.0789 & 6.6 & -2.2 & 6.6 & 0.3047 & 0.0197 & 1800 & 30.4 \\
58071.67853 & 0.63878 & 3.32 & 7.1014 & 6.6 & -18.2 & 6.6 & 0.3304 & 0.0274 & 1800 & 30.9 \\
58072.51203 & 0.63011 & 4.26 & 7.1227 & 8.5 & -34.7 & 8.5 & 0.2559 & 0.0242 & 1800 & 24.9 \\
58072.53371 & 0.63216 & 3.77 & 7.0956 & 7.5 & -13.4 & 7.5 & 0.2297 & 0.0221 & 1800 & 27.2 \\
58073.51729 & 0.63543 & 2.28 & 7.1075 & 4.6 & -12.7 & 4.6 & 0.2780 & 0.0116 & 1800 & 41.6 \\
58073.56164 & 0.64998 & 2.77 & 7.1045 & 5.5 & -0.2 & 5.5 & 0.2824 & 0.0190 & 1800 & 35.6 \\
58073.60189 & 0.63901 & 2.74 & 7.0926 & 5.5 & -16.7 & 5.5 & 0.2565 & 0.0202 & 1800 & 36.3 \\
58074.53118 & 0.63641 & 2.68 & 7.1044 & 5.4 & -10.9 & 5.4 & 0.2803 & 0.0153 & 1800 & 35.9 \\
58074.57702 & 0.64533 & 2.65 & 7.0932 & 5.3 & -14.0 & 5.3 & 0.2745 & 0.0152 & 1800 & 36.3 \\
58074.62708 & 0.63662 & 3.29 & 7.1002 & 6.6 & -1.4 & 6.6 & 0.2538 & 0.0225 & 1800 & 30.6 \\
58075.51759 & 0.64606 & 3.33 & 7.1055 & 6.7 & 17.5 & 6.7 & 0.2622 & 0.0196 & 1800 & 30.2 \\
58075.56099 & 0.64038 & 3.27 & 7.1078 & 6.5 & 3.8 & 6.5 & 0.2690 & 0.0199 & 1800 & 30.9 \\
58075.60824 & 0.64787 & 3.39 & 7.0843 & 6.8 & -18.8 & 6.8 & 0.2841 & 0.0193 & 1800 & 29.7 \\
58076.54917 & 0.65308 & 2.98 & 7.1028 & 6.0 & 6.9 & 6.0 & 0.2638 & 0.0177 & 1800 & 33.2 \\
58076.57087 & 0.65751 & 3.45 & 7.0690 & 6.9 & -1.4 & 6.9 & 0.2709 & 0.0218 & 1800 & 29.2 \\
58076.61758 & 0.64741 & 4.61 & 7.1043 & 9.2 & -34.8 & 9.2 & 0.2831 & 0.0328 & 1800 & 23.6 \\
58077.52776 & 0.64480 & 2.70 & 7.1019 & 5.4 & -16.7 & 5.4 & 0.2710 & 0.0167 & 1800 & 36.3 \\
58077.57256 & 0.64709 & 2.99 & 7.0781 & 6.0 & 8.6 & 6.0 & 0.2700 & 0.0193 & 1800 & 33.3 \\
58077.61695 & 0.65536 & 3.37 & 7.0696 & 6.7 & 3.2 & 6.7 & 0.2467 & 0.0233 & 1800 & 30.1 \\
58078.56898 & 0.64022 & 2.96 & 7.0962 & 5.9 & -10.3 & 5.9 & 0.2783 & 0.0233 & 1800 & 33.6 \\
58078.59137 & 0.63706 & 3.11 & 7.0994 & 6.2 & -6.2 & 6.2 & 0.2497 & 0.0219 & 1800 & 32.1 \\
58078.62745 & 0.64250 & 3.01 & 7.0851 & 6.0 & -5.9 & 6.0 & 0.2720 & 0.0199 & 1800 & 33.2 \\
58079.51543 & 0.63405 & 2.77 & 7.1093 & 5.5 & -1.6 & 5.5 & 0.2451 & 0.0144 & 1800 & 35.1 \\
58079.54796 & 0.62945 & 3.00 & 7.0837 & 6.0 & 17.2 & 6.0 & 0.2611 & 0.0221 & 1800 & 33.1 \\
58079.59431 & 0.63133 & 2.78 & 7.0932 & 5.6 & -0.7 & 5.6 & 0.2768 & 0.0145 & 1800 & 34.9 \\
58081.55658 & 0.64119 & 2.45 & 7.0684 & 4.9 & 0.9 & 4.9 & 0.2389 & 0.0155 & 1800 & 39.1 \\
58081.57821 & 0.63429 & 2.54 & 7.0772 & 5.1 & -0.3 & 5.1 & 0.2428 & 0.0165 & 1800 & 38.1 \\
58081.60094 & 0.63914 & 2.75 & 7.0767 & 5.5 & -19.9 & 5.5 & 0.2451 & 0.0184 & 1800 & 35.5 \\
58088.53242 & 0.64258 & 4.60 & 7.0893 & 9.2 & 8.0 & 9.2 & 0.2390 & 0.0261 & 1800 & 23.4 \\
58088.55272 & 0.64266 & 3.79 & 7.1093 & 7.6 & -2.4 & 7.6 & 0.2613 & 0.0218 & 1800 & 27.3 \\
58088.57529 & 0.64226 & 3.98 & 7.0745 & 8.0 & 2.7 & 8.0 & 0.2677 & 0.0232 & 1800 & 26.1 \\
58088.62166 & 0.64672 & 4.14 & 7.0986 & 8.3 & -12.8 & 8.3 & 0.2624 & 0.0272 & 1800 & 25.8 \\
58090.56233 & 0.62727 & 5.59 & 7.1181 & 11.2 & -21.0 & 11.2 & 0.2968 & 0.0357 & 1800 & 20.5 \\
58090.58262 & 0.62962 & 5.38 & 7.1014 & 10.8 & 4.0 & 10.8 & 0.2474 & 0.0354 & 1800 & 21.3 \\
58090.60496 & 0.62293 & 5.27 & 7.0521 & 10.5 & 0.4 & 10.5 & 0.2031 & 0.0378 & 1800 & 21.7 \\
58091.54531 & 0.63908 & 2.83 & 7.1068 & 5.7 & 0.2 & 5.7 & 0.2734 & 0.0189 & 1800 & 35.0 \\
58091.56736 & 0.64540 & 3.13 & 7.0864 & 6.3 & 6.8 & 6.3 & 0.2680 & 0.0220 & 1800 & 32.0 \\
58091.58933 & 0.65154 & 3.79 & 7.0775 & 7.6 & 15.2 & 7.6 & 0.2267 & 0.0256 & 1800 & 27.6 \\
58092.53743 & 0.65831 & 4.19 & 7.1000 & 8.4 & -10.7 & 8.4 & 0.2361 & 0.0243 & 1800 & 24.9 \\
58092.55954 & 0.65182 & 4.56 & 7.0921 & 9.1 & -24.4 & 9.1 & 0.2673 & 0.0275 & 1800 & 23.4 \\
58092.58185 & 0.64852 & 4.11 & 7.0815 & 8.2 & -1.6 & 8.2 & 0.2635 & 0.0257 & 1800 & 25.5 \\
58092.60390 & 0.64151 & 4.57 & 7.0793 & 9.1 & -13.1 & 9.1 & 0.2529 & 0.0304 & 1800 & 23.8 \\
58257.88259 & 0.63150 & 4.20 & 7.1000 & 8.4 & -19.1 & 8.4 & 0.2595 & 0.0195 & 1800 & 24.1 \\
58257.90337 & 0.63058 & 3.71 & 7.1019 & 7.4 & -4.1 & 7.4 & 0.2678 & 0.0162 & 1800 & 26.4 \\
58258.90198 & 0.63305 & 3.24 & 7.0838 & 6.5 & -18.3 & 6.5 & 0.2470 & 0.0133 & 1800 & 29.3 \\
58258.92463 & 0.63191 & 4.06 & 7.0784 & 8.1 & -2.6 & 8.1 & 0.2757 & 0.0180 & 1800 & 24.6 \\
58259.90731 & 0.64576 & 6.71 & 7.0633 & 13.4 & -5.4 & 13.4 & 0.1788 & 0.0321 & 1800 & 16.7 \\
58259.92809 & 0.63558 & 6.60 & 7.0586 & 13.2 & -8.4 & 13.2 & 0.2748 & 0.0310 & 1800 & 16.9 \\
58260.89447 & 0.63393 & 2.74 & 7.0810 & 5.5 & -8.5 & 5.5 & 0.2679 & 0.0110 & 1800 & 33.7 \\
58260.91546 & 0.63087 & 2.54 & 7.0739 & 5.1 & -7.2 & 5.1 & 0.2748 & 0.0100 & 1800 & 35.9 \\
58261.90506 & 0.63243 & 2.98 & 7.0733 & 6.0 & -16.8 & 6.0 & 0.2321 & 0.0116 & 1800 & 31.0 \\
58261.92647 & 0.62968 & 2.99 & 7.0719 & 6.0 & -15.6 & 6.0 & 0.2496 & 0.0118 & 1800 & 31.2 \\
58271.84868 & 0.62979 & 6.45 & 7.0761 & 12.9 & 0.4 & 12.9 & 0.2666 & 0.0311 & 1800 & 17.3 \\
58271.91011 & 0.63896 & 5.21 & 7.0613 & 10.4 & -30.7 & 10.4 & 0.2539 & 0.0237 & 1800 & 20.4 \\
58272.85336 & 0.63311 & 3.36 & 7.0823 & 6.7 & -2.4 & 6.7 & 0.2753 & 0.0139 & 1800 & 28.5 \\
58272.90937 & 0.63125 & 3.86 & 7.0780 & 7.7 & 6.0 & 7.7 & 0.2206 & 0.0155 & 1800 & 25.2 \\
58273.87346 & 0.63622 & 3.83 & 7.0829 & 7.7 & -15.2 & 7.7 & 0.2625 & 0.0164 & 1800 & 25.9 \\
58273.93527 & 0.63527 & 4.17 & 7.0835 & 8.3 & 0.6 & 8.3 & 0.2519 & 0.0184 & 1800 & 24.2 \\
58274.84530 & 0.62242 & 4.85 & 7.0869 & 9.7 & -11.2 & 9.7 & 0.1993 & 0.0226 & 1700 & 21.6 \\
58274.93420 & 0.63490 & 3.35 & 7.0902 & 6.7 & -1.0 & 6.7 & 0.2258 & 0.0152 & 1700 & 29.0 \\
58275.85645 & 0.62957 & 6.44 & 7.0635 & 12.9 & -4.4 & 12.9 & 0.2475 & 0.0282 & 1800 & 17.4 \\
58276.88598 & 0.63749 & 3.78 & 7.0675 & 7.6 & -10.8 & 7.6 & 0.2422 & 0.0174 & 1800 & 26.4 \\
58277.88791 & 0.63431 & 2.90 & 7.0675 & 5.8 & -3.9 & 5.8 & 0.2692 & 0.0120 & 1800 & 32.1 \\
58282.87925 & 0.63641 & 2.43 & 7.0800 & 4.9 & -15.1 & 4.9 & 0.2547 & 0.0096 & 1800 & 37.6 \\
58282.90241 & 0.64674 & 2.25 & 7.0796 & 4.5 & -18.2 & 4.5 & 0.2645 & 0.0094 & 1800 & 40.8 \\
58283.85144 & 0.63995 & 2.82 & 7.0887 & 5.6 & -1.7 & 5.6 & 0.2640 & 0.0108 & 1800 & 32.7 \\
58283.93521 & 0.63938 & 3.10 & 7.0884 & 6.2 & -15.9 & 6.2 & 0.2443 & 0.0137 & 1700 & 30.8 \\
58284.88839 & 0.63670 & 2.80 & 7.0835 & 5.6 & -17.0 & 5.6 & 0.2607 & 0.0121 & 1800 & 33.5 \\
58284.92110 & 0.63292 & 3.38 & 7.0843 & 6.8 & -12.9 & 6.8 & 0.2534 & 0.0159 & 1800 & 28.7 \\
58286.85685 & 0.63050 & 3.53 & 7.0764 & 7.1 & -4.5 & 7.1 & 0.2363 & 0.0153 & 1800 & 27.3 \\
58286.92281 & 0.63926 & 3.99 & 7.0794 & 8.0 & 7.7 & 8.0 & 0.2476 & 0.0186 & 1800 & 25.0 \\
58287.85433 & 0.64122 & 3.96 & 7.0579 & 7.9 & -18.7 & 7.9 & 0.2233 & 0.0179 & 1800 & 25.0 \\
58287.92379 & 0.64268 & 4.12 & 7.0489 & 8.2 & -4.3 & 8.2 & 0.2696 & 0.0199 & 1800 & 24.4 \\
58288.87659 & 0.63222 & 3.30 & 7.0591 & 6.6 & -4.4 & 6.6 & 0.2451 & 0.0143 & 1800 & 29.0 \\
58288.89757 & 0.63579 & 3.45 & 7.0922 & 6.9 & 1.1 & 6.9 & 0.2824 & 0.0157 & 1800 & 28.1 \\
58289.87215 & 0.63318 & 2.47 & 7.0769 & 4.9 & -14.6 & 4.9 & 0.2580 & 0.0099 & 1800 & 36.8 \\
58289.89334 & 0.63750 & 2.42 & 7.0770 & 4.8 & -9.2 & 4.8 & 0.2347 & 0.0101 & 1800 & 37.9 \\
58290.77814 & 0.64916 & 3.98 & 7.0836 & 8.0 & 7.9 & 8.0 & 0.2668 & 0.0185 & 1800 & 25.3 \\
58290.79955 & 0.64048 & 3.29 & 7.0895 & 6.6 & -10.1 & 6.6 & 0.2589 & 0.0141 & 1800 & 29.0 \\
58291.89620 & 0.63694 & 3.54 & 7.0835 & 7.1 & -14.5 & 7.1 & 0.2840 & 0.0170 & 1800 & 27.7 \\
58291.91677 & 0.64216 & 2.94 & 7.0964 & 5.9 & 5.3 & 5.9 & 0.2561 & 0.0138 & 1800 & 32.4 \\
58292.83057 & 0.64020 & 3.59 & 7.0777 & 7.2 & -10.0 & 7.2 & 0.2536 & 0.0145 & 1800 & 26.7 \\
58292.85135 & 0.64129 & 3.59 & 7.0688 & 7.2 & -5.1 & 7.2 & 0.2780 & 0.0145 & 1800 & 26.8 \\
58295.83519 & 0.63730 & 2.54 & 7.0811 & 5.1 & -10.3 & 5.1 & 0.2655 & 0.0100 & 1800 & 36.2 \\
58295.85638 & 0.63307 & 2.58 & 7.0843 & 5.2 & -1.5 & 5.2 & 0.2632 & 0.0108 & 1800 & 36.0 \\
58298.78493 & 0.64344 & 3.00 & 7.0801 & 6.0 & -9.0 & 6.0 & 0.2741 & 0.0119 & 1800 & 31.5 \\
58298.85235 & 0.65183 & 3.37 & 7.0844 & 6.7 & -2.6 & 6.7 & 0.2603 & 0.0137 & 1800 & 28.6 \\
58306.81818 & 0.63824 & 5.89 & 7.0798 & 11.8 & -11.8 & 11.8 & 0.2418 & 0.0280 & 1800 & 18.5 \\
58306.85366 & 0.64448 & 4.81 & 7.0616 & 9.6 & -21.1 & 9.6 & 0.2636 & 0.0235 & 1800 & 21.7 \\
58307.83986 & 0.64326 & 2.79 & 7.0759 & 5.6 & -7.6 & 5.6 & 0.2303 & 0.0121 & 1800 & 33.8 \\
58307.89270 & 0.64364 & 3.85 & 7.0774 & 7.7 & -9.3 & 7.7 & 0.1984 & 0.0192 & 1800 & 26.1 \\
58309.84695 & 0.64134 & 3.70 & 7.0563 & 7.4 & -8.8 & 7.4 & 0.2404 & 0.0179 & 1800 & 26.7 \\
58309.86856 & 0.64961 & 3.72 & 7.0774 & 7.4 & 3.0 & 7.4 & 0.2505 & 0.0184 & 1800 & 26.7 \\
58310.83607 & 0.63253 & 3.91 & 7.0765 & 7.8 & -27.7 & 7.8 & 0.2574 & 0.0176 & 1800 & 25.4 \\
58310.93220 & 0.64379 & 2.73 & 7.0533 & 5.5 & -2.1 & 5.5 & 0.2307 & 0.0130 & 1800 & 34.7 \\
58311.85812 & 0.63401 & 3.70 & 7.0329 & 7.4 & -26.6 & 7.4 & 0.2190 & 0.0174 & 1800 & 26.5 \\
58311.87890 & 0.64133 & 3.41 & 7.0551 & 6.8 & -12.2 & 6.8 & 0.2477 & 0.0158 & 1800 & 28.5 \\
58312.78337 & 0.62907 & 2.91 & 7.0687 & 5.8 & -5.6 & 5.8 & 0.2445 & 0.0120 & 1800 & 32.1 \\
58312.80435 & 0.62907 & 2.92 & 7.0671 & 5.8 & -24.3 & 5.8 & 0.2263 & 0.0124 & 1800 & 32.2 \\
58334.68745 & 0.65190 & 3.25 & 7.0722 & 6.5 & 11.4 & 6.5 & 0.2911 & 0.0142 & 1800 & 29.2 \\
58335.70929 & 0.62999 & 5.11 & 7.0627 & 10.2 & -6.3 & 10.2 & 0.2028 & 0.0245 & 1800 & 20.7 \\
58335.77584 & 0.64210 & 3.39 & 7.0884 & 6.8 & -9.0 & 6.8 & 0.2602 & 0.0155 & 1800 & 28.6 \\
58336.71081 & 0.63754 & 2.98 & 7.0712 & 6.0 & -16.5 & 6.0 & 0.2630 & 0.0122 & 1800 & 31.3 \\
58336.77381 & 0.63792 & 3.46 & 7.0805 & 6.9 & 0.1 & 6.9 & 0.2323 & 0.0161 & 1800 & 28.0 \\
58336.88642 & 0.63870 & 3.63 & 7.0751 & 7.3 & 9.7 & 7.3 & 0.2557 & 0.0178 & 1800 & 27.3 \\
58338.89994 & 0.62874 & 4.11 & 7.0753 & 8.2 & 3.5 & 8.2 & 0.2348 & 0.0234 & 1800 & 24.8 \\
58339.70411 & 0.63470 & 2.64 & 7.0711 & 5.3 & -9.0 & 5.3 & 0.2376 & 0.0104 & 1800 & 34.5 \\
58339.75402 & 0.62799 & 2.78 & 7.0536 & 5.6 & -4.8 & 5.6 & 0.2352 & 0.0120 & 1800 & 33.5 \\
58341.68094 & 0.64922 & 4.38 & 7.0387 & 8.8 & -15.6 & 8.8 & 0.2446 & 0.0195 & 1800 & 22.9 \\
58341.80931 & 0.64243 & 4.61 & 7.0299 & 9.2 & -9.3 & 9.2 & 0.2242 & 0.0205 & 1800 & 21.7 \\
58341.89764 & 0.64930 & 3.57 & 7.0545 & 7.1 & -13.3 & 7.1 & 0.2043 & 0.0194 & 1800 & 27.9 \\
58342.65721 & 0.64728 & 4.69 & 7.0674 & 9.4 & 3.0 & 9.4 & 0.2220 & 0.0203 & 1800 & 21.6 \\
58342.90281 & 0.63838 & 3.42 & 7.0514 & 6.8 & -7.9 & 6.8 & 0.2171 & 0.0194 & 1800 & 29.1 \\
58343.68619 & 0.64121 & 7.02 & 7.0560 & 14.0 & -0.7 & 14.0 & 0.2056 & 0.0375 & 1800 & 16.1 \\
58343.75359 & 0.61768 & 8.24 & 7.0511 & 16.5 & 8.5 & 16.5 & 0.2560 & 0.0486 & 1800 & 14.5 \\
58344.66308 & 0.63558 & 2.62 & 7.0574 & 5.2 & -19.6 & 5.2 & 0.2504 & 0.0098 & 1800 & 34.5 \\
58344.73397 & 0.63875 & 3.07 & 7.0678 & 6.1 & -5.5 & 6.1 & 0.2181 & 0.0136 & 1800 & 30.9 \\
58344.80023 & 0.62905 & 4.15 & 7.0413 & 8.3 & -4.6 & 8.3 & 0.2748 & 0.0215 & 1800 & 24.3 \\
58346.65788 & 0.64317 & 2.43 & 7.0551 & 4.9 & -11.1 & 4.9 & 0.2550 & 0.0089 & 1800 & 36.9 \\
58346.72846 & 0.64000 & 2.52 & 7.0649 & 5.0 & -6.5 & 5.0 & 0.2488 & 0.0113 & 1800 & 36.7 \\
58347.66173 & 0.64481 & 2.19 & 7.0706 & 4.4 & -12.6 & 4.4 & 0.2375 & 0.0077 & 1800 & 40.8 \\
58347.80065 & 0.64974 & 2.58 & 7.0596 & 5.2 & -11.0 & 5.2 & 0.2336 & 0.0126 & 1800 & 36.4 \\
58349.67671 & 0.64287 & 3.52 & 7.0571 & 7.0 & 5.3 & 7.0 & 0.2316 & 0.0153 & 1800 & 27.5 \\
58349.75317 & 0.64453 & 3.16 & 7.0692 & 6.3 & -22.1 & 6.3 & 0.2476 & 0.0143 & 1800 & 30.2 \\
58350.61713 & 0.64621 & 4.01 & 7.0618 & 8.0 & -15.4 & 8.0 & 0.2368 & 0.0174 & 1800 & 24.8 \\
58350.69100 & 0.64525 & 2.55 & 7.0724 & 5.1 & -15.4 & 5.1 & 0.2278 & 0.0103 & 1800 & 36.2 \\
58350.76720 & 0.64174 & 2.91 & 7.0616 & 5.8 & -16.6 & 5.8 & 0.2376 & 0.0135 & 1800 & 32.6 \\
58351.61917 & 0.64931 & 3.57 & 7.0676 & 7.1 & -14.4 & 7.1 & 0.2555 & 0.0154 & 1800 & 27.3 \\
58351.85181 & 0.64384 & 3.26 & 7.0692 & 6.5 & -22.7 & 6.5 & 0.2263 & 0.0189 & 1800 & 30.2 \\
58352.66763 & 0.63827 & 2.67 & 7.0782 & 5.3 & -10.9 & 5.3 & 0.2519 & 0.0106 & 1800 & 34.6 \\
58352.71171 & 0.64317 & 2.76 & 7.0808 & 5.5 & -4.0 & 5.5 & 0.2413 & 0.0124 & 1800 & 34.2 \\
58352.76784 & 0.64613 & 2.39 & 7.0593 & 4.8 & -8.1 & 4.8 & 0.2517 & 0.0113 & 1800 & 39.2 \\
58362.80959 & 0.63901 & 11.82 & 7.0614 & 23.6 & 6.8 & 23.6 & 0.1411 & 0.0709 & 1800 & 10.8 \\
58363.61078 & 0.63047 & 3.49 & 7.0380 & 7.0 & -14.9 & 7.0 & 0.2109 & 0.0151 & 1800 & 27.5 \\
58363.71811 & 0.63923 & 2.75 & 7.0474 & 5.5 & -25.5 & 5.5 & 0.2297 & 0.0136 & 1800 & 34.3 \\
58363.81133 & 0.63093 & 3.82 & 7.0228 & 7.6 & -11.2 & 7.6 & 0.2731 & 0.0224 & 1800 & 26.4 \\
58365.78692 & 0.63255 & 3.59 & 7.0524 & 7.2 & -17.6 & 7.2 & 0.2406 & 0.0199 & 1800 & 27.9 \\
58365.85687 & 0.64228 & 3.73 & 7.0503 & 7.5 & 4.8 & 7.5 & 0.2558 & 0.0222 & 1800 & 27.1 \\
\end{longtable}

\newpage

\begin{landscape}
\begin{longtable}{lccc}
\caption{\label{MCMCprior} List of parameters used in the analysis. The respective priors are provided together with the posteriors for the Dartmouth and PARSEC stellar evolution tracks. The posterior values represent the median and 68.3\% credible interval. Derived values that might be useful for follow-up work are also reported.}\\
\hline
Parameter & Prior & \multicolumn{2}{c}{Posterior}\\
 &  & Dartmouth & PARSEC\\
 &  & (adopted) & \\
\hline
\endfirsthead
\multicolumn{4}{l}{{\bfseries \tablename\ \thetable{} -- continued from previous page}} \\
\hline
Parameter & Prior & \multicolumn{2}{c}{Posterior}\\
 &  & Dartmouth & PARSEC\\
\hline
\endhead
\multicolumn{4}{l}{{Continued on next page}} \\ 
\hline
\endfoot
\hline
\multicolumn{4}{l}{Notes:}\\
\multicolumn{4}{l}{$\bullet$ $\mathcal{N}(\mu,\sigma^{2})$: Normal distribution with mean $\mu$ and width $\sigma^{2}$}\\
\multicolumn{4}{l}{$\bullet$ $\mathcal{U}(a,b)$: Uniform distribution between $a$ and $b$}\\
\multicolumn{4}{l}{$\bullet$ $\mathcal{S}(a,b)$: Sine distribution between $a$ and $b$}\\
\multicolumn{4}{l}{$\bullet$ $\mathcal{T}(\mu,\sigma^{2},a,b)$: Truncated normal distribution with mean $\mu$ and width $\sigma^{2}$, between $a$ and $b$}\\
\endlastfoot
\\
\multicolumn{4}{l}{\it Stellar Parameters}\\
\\
Effective temperature \teff\ [K] & $\mathcal{N}(5350.0, 80.0)$ & $5356.3^{_{+41.8}}_{^{-13.1}}$ & $5424.1^{_{+21.0}}_{^{-32.0}}$ \\
Surface gravity \logg\ [cgs] & $\mathcal{N}(4.52,0.15)$ & $4.54^{_{+0.02}}_{^{-0.04}}$ & $4.56^{_{+0.01}}_{^{-0.02}}$ \\
Metallicity [M/H] [dex] &  $\mathcal{N}(0.15,0.04)$ & $0.14^{_{+0.04}}_{^{-0.04}}$ & $0.12^{_{+0.04}}_{^{-0.03}}$ \\
Distance to Earth $D$ [pc] &  $\mathcal{N}(201.66,6.38)$ & $208.9^{_{+6.5}}_{^{-4.9}}$ & $204.1^{_{+4.3}}_{^{-1.0}}$ \\
Interstellar extinction $E(B-V)$ [mag] &  $\mathcal{U}(0.0,1.0)$ & $0.007^{_{+0.010}}_{^{-0.005}}$ & $0.019^{_{+0.012}}_{^{-0.012}}$ \\
Systemic radial velocity $\gamma$ [\kms] & $\mathcal{U}(-10.0,10.0)$ & $0.6386^{_{+0.0023}}_{^{-0.0026}}$ & $0.6386^{_{+0.0029}}_{^{-0.0026}}$ \\
Linear limb-darkening coefficient $u_{a}$ & (derived) & $0.4962^{_{+0.0051}}_{^{-0.0101}}$ & $0.4795^{_{+0.0097}}_{^{-0.0060}}$ \\
Quadratic limb-darkening coefficient $u_{b}$ & (derived) & $0.2055^{_{+0.0067}}_{^{-0.0032}}$ & $0.2165^{_{+0.0039}}_{^{-0.0062}}$ \\
Stellar density $\rho_{\star}/\rho_{\astrosun}$ & (derived) & $1.464^{_{+0.097}}_{^{-0.186}}$ & $1.570^{_{+0.059}}_{^{-0.092}}$ \\
Stellar mass M$_{\star}$\ [\Msun] & (derived) & $0.935^{_{+0.019}}_{^{-0.023}}$ & $0.939^{_{+0.011}}_{^{-0.020}}$ \\
Stellar radius R$_{\star}$\ [\Rsun] & (derived) & $0.863^{_{+0.031}}_{^{-0.018}}$ & $0.841^{_{+0.016}}_{^{-0.008}}$ \\
Stellar age $\tau$\ [Gyr] & (derived) & $2.8^{_{+3.8}}_{^{-1.7}}$ & $1.2^{_{+1.7}}_{^{-0.8}}$ \\
&&& \\
\hline
\\
\multicolumn{4}{l}{\it Planet b Parameters}\\
\\
Orbital Period $P_{b}$ [d] &  $\mathcal{N}(2.35322,0.01)$ & $2.35309^{_{+0.00022}}_{^{-0.00022}}$ & $2.35310^{_{+0.00021}}_{^{-0.00022}}$ \\
Transit epoch $T_{0,b}$ [BJD - 2450000] &  $\mathcal{N}(7773.317,0.001)$ & $7773.31683^{_{+0.00091}}_{^{-0.00090}}$ & $7773.31688^{_{+0.00096}}_{^{-0.00092}}$ \\
Radial velocity semi-amplitude $K_{b}$ [\kms] & $\mathcal{U}(0.0,0.1)$ & $0.00156^{_{+0.00053}}_{^{-0.00053}}$ & $0.00154^{_{+0.00050}}_{^{-0.00054}}$ \\
Orbital inclination $i_{b}$ [$^{\circ}$] & $\mathcal{S}(70.0,90.0)$ & $87.2^{_{+1.2}}_{^{-1.0}}$ & $87.8^{_{+1.5}}_{^{-0.6}}$ \\
Planet-to-star radius ratio $k_{b}$ & $\mathcal{U}(0.0,1.0)$ & $0.01600^{_{+0.00083}}_{^{-0.00076}}$ & $0.01581^{_{+0.00063}}_{^{-0.00060}}$ \\
Orbital eccentricity $e_{b}$ & $\mathcal{T}(0.0,0.083,0.0,1.0)$ & $0.048^{_{+0.054}}_{^{-0.033}}$ & $0.053^{_{+0.059}}_{^{-0.038}}$ \\
Argument of periastron $\omega_{b}$ [\degr] & $\mathcal{U}(0.0,360.0)$ & $244^{_{+32}}_{^{-97}}$ & $194^{_{+64}}_{^{-100}}$ \\
System scale $a_{b}/R_{\star}$ & (derived) & $8.5^{_{+0.2}}_{^{-0.4}}$ & $8.7^{_{+0.1}}_{^{-0.2}}$ \\
Impact parameter $b_{b}$ & (derived) & $0.419^{_{+0.164}}_{^{-0.181}}$ & $0.332^{_{+0.095}}_{^{-0.229}}$ \\
Transit duration T$_{14,b}$ [h] & (derived) & $2.01^{_{+0.12}}_{^{-0.12}}$ & $2.01^{_{+0.09}}_{^{-0.08}}$ \\
Semi-major axis $a_{b}$ [AU] & (derived) & $0.03385^{_{+0.00023}}_{^{-0.00029}}$ & $0.03390^{_{+0.00013}}_{^{-0.00024}}$ \\
Planet mass M$_{b}$ [\Mearth] & (derived) & $3.10^{_{+1.05}}_{^{-1.05}}$ & $3.06^{_{+1.01}}_{^{-1.05}}$ \\
Planet radius R$_{b}$ [\Rearth] & (derived) & $1.510^{_{+0.110}}_{^{-0.084}}$ & $1.454^{_{+0.063}}_{^{-0.058}}$ \\
Planet bulk density $\rho_{b}$ [\gcm3] & (derived) & $4.9^{_{+2.0}}_{^{-1.7}}$ & $5.4^{_{+2.1}}_{^{-1.9}}$ \\
&&& \\
\hline
\\
\multicolumn{4}{l}{\it Planet c Parameters}\\
\\
Orbital Period $P_{c}$ [d] &  $\mathcal{N}(3.55987,0.01)$ & $3.56004^{_{+0.00012}}_{^{-0.00011}}$ & $3.56004^{_{+0.00010}}_{^{-0.00011}}$ \\
Transit epoch $T_{0,c}$ [BJD - 2450000] &  $\mathcal{N}(7740.3223,0.001)$ & $7740.32182^{_{+0.00087}}_{^{-0.00089}}$ & $7740.32191^{_{+0.00085}}_{^{-0.00088}}$ \\
Radial velocity semi-amplitude $K_{c}$ [\kms] & $\mathcal{U}(0.0,0.1)$ & $0.00278^{_{+0.00050}}_{^{-0.00055}}$ & $0.00277^{_{+0.00056}}_{^{-0.00050}}$ \\
Orbital inclination $i_{c}$ [$^{\circ}$] & $\mathcal{S}(70.0,90.0)$ & $88.1^{_{+0.7}}_{^{-0.7}}$ & $88.4^{_{+0.8}}_{^{-0.7}}$ \\
Planet-to-star radius ratio $k_{c}$ & $\mathcal{U}(0.0,1.0)$ & $0.02439^{_{+0.00065}}_{^{-0.00059}}$ & $0.02428^{_{+0.00063}}_{^{-0.00057}}$ \\
Orbital eccentricity $e_{c}$ & $\mathcal{T}(0.0,0.083,0.0,1.0)$ & $0.045^{_{+0.051}}_{^{-0.032}}$ & $0.036^{_{+0.049}}_{^{-0.025}}$ \\
Argument of periastron $\omega_{c}$ [\degr] & $\mathcal{U}(0.0,360.0)$ & $200^{_{+60}}_{^{-27}}$ & $238^{_{+23}}_{^{-95}}$ \\
System scale $a_{c}/R_{\star}$ & (derived) & $11.1^{_{+0.2}}_{^{-0.5}}$ & $11.4^{_{+0.1}}_{^{-0.2}}$ \\
Impact parameter $b_{c}$ & (derived) & $0.369^{_{+0.130}}_{^{-0.136}}$ & $0.316^{_{+0.145}}_{^{-0.162}}$ \\
Transit duration T$_{14,c}$ [h] & (derived) & $2.38^{_{+0.05}}_{^{-0.06}}$ & $2.38^{_{+0.07}}_{^{-0.06}}$ \\
Semi-major axis $a_{c}$ [AU] & (derived) & $0.04461^{_{+0.00030}}_{^{-0.00038}}$ & $0.04468^{_{+0.00018}}_{^{-0.00031}}$ \\
Planet mass M$_{c}$ [\Mearth] & (derived) & $6.31^{_{+1.13}}_{^{-1.23}}$ & $6.30^{_{+1.28}}_{^{-1.14}}$ \\
Planet radius R$_{c}$ [\Rearth] & (derived) & $2.299^{_{+0.120}}_{^{-0.087}}$ & $2.231^{_{+0.081}}_{^{-0.062}}$ \\
Planet bulk density $\rho_{c}$ [\gcm3] & (derived) & $2.8^{_{+0.7}}_{^{-0.6}}$ & $3.1^{_{+0.7}}_{^{-0.6}}$ \\
&&& \\
\hline
\\
\multicolumn{4}{l}{\it Planet d Parameters}\\
\\
Orbital Period $P_{d}$ [d] &  $\mathcal{N}(5.40478,0.01)$ & $5.40479^{_{+0.00021}}_{^{-0.00021}}$ & $5.40481^{_{+0.00020}}_{^{-0.00021}}$ \\
Transit epoch $T_{0,d}$ [BJD - 2450000] &  $\mathcal{N}(7743.1607,0.001)$ & $7743.15989^{_{+0.00091}}_{^{-0.00093}}$ & $7743.15987^{_{+0.00092}}_{^{-0.00094}}$ \\
Radial velocity semi-amplitude $K_{d}$ [\kms] & $\mathcal{U}(0.0,0.1)$ & $0.00303\pm0.00052$ & $0.00303\pm0.00051$ \\
Orbital inclination $i_{d}$ [$^{\circ}$] & $\mathcal{S}(70.0,90.0)$ & $89.0^{_{+0.6}}_{^{-0.6}}$ & $89.2^{_{+0.6}}_{^{-0.5}}$ \\
Planet-to-star radius ratio $k_{d}$ & $\mathcal{U}(0.0,1.0)$ & $0.02532^{_{+0.00067}}_{^{-0.00063}}$ & $0.02528^{_{+0.00064}}_{^{-0.00060}}$ \\
Orbital eccentricity $e_{d}$ & $\mathcal{T}(0.0,0.083,0.0,1.0)$ & $0.043^{_{+0.041}}_{^{-0.030}}$ & $0.033^{_{+0.032}}_{^{-0.024}}$ \\
Argument of periastron $\omega_{d}$ [\degr] & $\mathcal{U}(0.0,360.0)$ & $83^{_{+82}}_{^{-38}}$ & $79^{_{+101}}_{^{-25}}$ \\
System scale $a_{d}/R_{\star}$ & (derived) & $14.7^{_{+0.3}}_{^{-0.7}}$ & $15.1^{_{+0.2}}_{^{-0.3}}$ \\
Impact parameter $b_{d}$ & (derived) & $0.241^{_{+0.159}}_{^{-0.152}}$ & $0.199^{_{+0.120}}_{^{-0.146}}$ \\
Transit duration T$_{14,d}$ [h] & (derived) & $2.70^{_{+0.07}}_{^{-0.07}}$ & $2.68^{_{+0.09}}_{^{-0.08}}$ \\
Semi-major axis $a_{d}$ [AU] & (derived) & $0.05893^{_{+0.00040}}_{^{-0.00050}}$ & $0.05902^{_{+0.00023}}_{^{-0.00041}}$ \\
Planet mass M$_{d}$ [\Mearth] & (derived) & $7.92^{_{+1.39}}_{^{-1.35}}$ & $7.94^{_{+1.36}}_{^{-1.33}}$ \\
Planet radius R$_{d}$ [\Rearth] & (derived) & $2.390^{_{+0.104}}_{^{-0.084}}$ & $2.328^{_{+0.066}}_{^{-0.061}}$ \\
Planet bulk density $\rho_{d}$ [\gcm3] & (derived) & $3.1^{_{+0.7}}_{^{-0.6}}$ & $3.5^{_{+0.6}}_{^{-0.6}}$ \\
&&& \\
\hline
\\
\multicolumn{4}{l}{\it Planet e Parameters}\\
\\
Orbital Period $P_{e}$ [d] &  $\mathcal{N}(8.26144,0.01)$ & $8.26146^{_{+0.00021}}_{^{-0.00022}}$ & $8.26147^{_{+0.00022}}_{^{-0.00022}}$ \\
Transit epoch $T_{0,e}$ [BJD - 2450000] &  $\mathcal{N}(7740.6451,0.001)$ & $7740.64558^{_{+0.00085}}_{^{-0.00081}}$ & $7740.64556^{_{+0.00085}}_{^{-0.00089}}$ \\
Radial velocity semi-amplitude $K_{e}$ [\kms] & $\mathcal{U}(0.0,0.1)$ & $0.00431^{_{+0.00067}}_{^{-0.00067}}$ & $0.00426^{_{+0.00062}}_{^{-0.00063}}$ \\
Orbital inclination $i_{e}$ [$^{\circ}$] & $\mathcal{S}(70.0,90.0)$ & $88.6^{_{+0.3}}_{^{-0.3}}$ & $88.8^{_{+0.4}}_{^{-0.2}}$ \\
Planet-to-star radius ratio $k_{e}$ & $\mathcal{U}(0.0,1.0)$ & $0.03599^{_{+0.00081}}_{^{-0.00073}}$ & $0.03575^{_{+0.00074}}_{^{-0.00075}}$ \\
Orbital eccentricity $e_{e}$ & $\mathcal{T}(0.0,0.083,0.0,1.0)$ & $0.077^{_{+0.048}}_{^{-0.049}}$ & $0.087^{_{+0.046}}_{^{-0.050}}$ \\
Argument of periastron $\omega_{e}$ [\degr] & $\mathcal{U}(0.0,360.0)$ & $180^{_{+40}}_{^{-44}}$ & $173^{_{+58}}_{^{-29}}$ \\
System scale $a_{e}/R_{\star}$ & (derived) & $19.5^{_{+0.4}}_{^{-0.9}}$ & $20.0^{_{+0.2}}_{^{-0.4}}$ \\
Impact parameter $b_{e}$ & (derived) & $0.468^{_{+0.096}}_{^{-0.126}}$ & $0.418^{_{+0.095}}_{^{-0.140}}$ \\
Transit duration T$_{14,e}$ [h] & (derived) & $2.97\pm0.05$ & $2.97\pm0.05$ \\
Semi-major axis $a_{e}$ [AU] & (derived) & $0.07820^{_{+0.00053}}_{^{-0.00066}}$ & $0.07832^{_{+0.00031}}_{^{-0.00055}}$ \\
Planet mass M$_{e}$ [\Mearth] & (derived) & $12.97^{_{+1.98}}_{^{-1.99}}$ & $12.79^{_{+1.88}}_{^{-1.91}}$ \\
Planet radius R$_{e}$ [\Rearth] & (derived) & $3.390^{_{+0.156}}_{^{-0.110}}$ & $3.290^{_{+0.087}}_{^{-0.084}}$ \\
Planet bulk density $\rho_{e}$ [\gcm3] & (derived) & $1.8^{_{+0.4}}_{^{-0.3}}$ & $2.0^{_{+0.3}}_{^{-0.3}}$ \\
&&& \\
\hline
\\
\multicolumn{4}{l}{\it Planet f Parameters}\\
\\
Orbital Period $P_{f}$ [d] &  $\mathcal{N}(12.75759,0.01)$ & $12.75758^{_{+0.00050}}_{^{-0.00048}}$ & $12.75758^{_{+0.00051}}_{^{-0.00046}}$ \\
Transit epoch $T_{0,f}$ [BJD - 2450000] &  $\mathcal{N}(7738.7019,0.001)$ & $7738.70235^{_{+0.00093}}_{^{-0.00092}}$ & $7738.70227^{_{+0.00094}}_{^{-0.00094}}$ \\
Radial velocity semi-amplitude $K_{f}$ [\kms] & $\mathcal{U}(0.0,0.1)$ & $0.00047^{_{+0.00061}}_{^{-0.00034}}$ & $0.00042^{_{+0.00056}}_{^{-0.00030}}$ \\
Orbital inclination $i_{f}$ [$^{\circ}$] & $\mathcal{S}(70.0,90.0)$ & $88.8^{_{+0.2}}_{^{-0.2}}$ & $88.9^{_{+0.2}}_{^{-0.1}}$ \\
Planet-to-star radius ratio $k_{f}$ & $\mathcal{U}(0.0,1.0)$ & $0.03085^{_{+0.00088}}_{^{-0.00089}}$ & $0.03049^{_{+0.00083}}_{^{-0.00083}}$ \\
Orbital eccentricity $e_{f}$ & $\mathcal{T}(0.0,0.083,0.0,1.0)$ & $0.062^{_{+0.064}}_{^{-0.043}}$ & $0.056^{_{+0.060}}_{^{-0.039}}$ \\
Argument of periastron $\omega_{f}$ [\degr] & $\mathcal{U}(0.0,360.0)$ & $144^{_{+117}}_{^{-80}}$ & $113^{_{+165}}_{^{-54}}$ \\
System scale $a_{f}/R_{\star}$ & (derived) & $26.1^{_{+0.6}}_{^{-1.2}}$ & $26.7^{_{+0.3}}_{^{-0.5}}$ \\
Impact parameter $b_{f}$ & (derived) & $0.560^{_{+0.079}}_{^{-0.112}}$ & $0.508^{_{+0.079}}_{^{-0.118}}$ \\
Transit duration T$_{14,f}$ [h] & (derived) & $3.20\pm0.08$ & $3.20\pm0.08$ \\
Semi-major axis $a_{f}$ [AU] & (derived) & $0.10447^{_{+0.00070}}_{^{-0.00088}}$ & $0.10463^{_{+0.00041}}_{^{-0.00073}}$ \\
Planet mass M$_{f}$ [\Mearth] & (derived) & $1.63^{_{+2.12}}_{^{-1.18}}$ & $1.46^{_{+1.98}}_{^{-1.05}}$ \\
Planet radius R$_{f}$ [\Rearth] & (derived) & $2.904^{_{+0.164}}_{^{-0.111}}$ & $2.803^{_{+0.102}}_{^{-0.090}}$ \\
Planet bulk density $\rho_{f}$ [\gcm3] & (derived) & $0.4^{_{+0.5}}_{^{-0.3}}$ & $0.4^{_{+0.5}}_{^{-0.3}}$ \\
&&& \\
\hline
\\
\multicolumn{4}{l}{\it Planet g Parameters}\\
\\
Orbital Period $P_{g}$ [d] &  $\mathcal{N}(41.97,0.1)$ & $41.96797^{_{+0.00843}}_{^{-0.00725}}$ & $41.96841^{_{+0.00928}}_{^{-0.00806}}$ \\
Transit epoch $T_{0,g}$ [BJD - 2450000] &  $\mathcal{N}(7773.76,2457773.93)$ & $7773.86156^{_{+0.01863}}_{^{-0.03219}}$ & $7773.87218^{_{+0.01034}}_{^{-0.01541}}$ \\
Radial velocity semi-amplitude $K_{g}$ [\kms] & $\mathcal{U}(0.0,1.0)$ & $0.00083^{_{+0.00102}}_{^{-0.00058}}$ & $0.00067^{_{+0.00078}}_{^{-0.00047}}$ \\
Orbital inclination $i_{g}$ [$^{\circ}$] & $\mathcal{S}(70.0,90.0)$ & $89.4^{_{+0.4}}_{^{-0.3}}$ & $89.7^{_{+0.2}}_{^{-0.3}}$ \\
Planet-to-star radius ratio $k_{g}$ & $\mathcal{U}(0.0,1.0)$ & $0.03187^{_{+0.00295}}_{^{-0.00250}}$ & $0.03084^{_{+0.00200}}_{^{-0.00212}}$ \\
Orbital eccentricity $e_{g}$ & $\mathcal{T}(0.0,0.083,0.0,1.0)$ & $0.059^{_{+0.063}}_{^{-0.040}}$ & $0.052^{_{+0.059}}_{^{-0.036}}$ \\
Argument of periastron $\omega_{g}$ [\degr] & $\mathcal{U}(0.0,360.0)$ & $164^{_{+92}}_{^{-72}}$ & $157^{_{+105}}_{^{-96}}$ \\
System scale $a_{g}/R_{\star}$ & (derived) & $57.7^{_{+1.2}}_{^{-2.6}}$ & $59.1^{_{+0.7}}_{^{-1.2}}$ \\
Impact parameter $b_{g}$ & (derived) & $0.558^{_{+0.284}}_{^{-0.356}}$ & $0.291^{_{+0.256}}_{^{-0.194}}$ \\
Transit duration T$_{14,g}$ [h] & (derived) & $4.75^{_{+0.80}}_{^{-1.42}}$ & $5.25^{_{+0.35}}_{^{-0.66}}$ \\
Semi-major axis $a_{g}$ [AU] & (derived) & $0.23109^{_{+0.00154}}_{^{-0.00196}}$ & $0.23144^{_{+0.00091}}_{^{-0.00162}}$ \\
Planet mass M$_{g}$ [\Mearth] & (derived) & $4.32^{_{+5.26}}_{^{-3.03}}$ & $3.48^{_{+4.05}}_{^{-2.41}}$ \\
Planet radius R$_{g}$ [\Rearth] & (derived) & $3.013^{_{+0.303}}_{^{-0.251}}$ & $2.837^{_{+0.196}}_{^{-0.197}}$ \\
Planet bulk density $\rho_{g}$ [\gcm3] & (derived) & $0.9^{_{+1.1}}_{^{-0.6}}$ & $0.8^{_{+1.0}}_{^{-0.6}}$ \\
&&& \\
\hline
&&& \\
\multicolumn{4}{l}{\it Instrument-related Parameters}\\
&&& \\
HARPS jitter $\sigma_{j,~\rm RV}$ [\kms] & $\mathcal{U}(0.0,0.1)$ & $0.00315^{_{+0.00043}}_{^{-0.00039}}$ & $0.00316^{_{+0.00041}}_{^{-0.00040}}$ \\
\textit{K2} contamination [\%] & $\mathcal{T}(0.0,0.005,0.0,1.0)$ & $0.004^{_{+0.004}}_{^{-0.003}}$ & $0.003^{_{+0.004}}_{^{-0.002}}$ \\
\textit{K2} jitter $\sigma_{j,~\textit{K2}}$ [ppm] & $\mathcal{U}(0.0, 10^5)$ & $185.8\pm2.6$ & $186.0\pm2.6$ \\
\textit{K2} out-of-transit flux & $\mathcal{U}(0.99,1.01)$ & $1.0000056^{_{+0.0000039}}_{^{-0.0000037}}$ & $1.0000057^{_{+0.0000038}}_{^{-0.0000036}}$ \\
SED jitter [mag]  & $\mathcal{U}(0.0,0.1)$ & $0.019^{_{+0.018}}_{^{-0.012}}$ & $0.015^{_{+0.016}}_{^{-0.011}}$ \\
&&& \\
\hline
\\
\multicolumn{4}{l}{\it Gaussian Process Hyperparameters}\\
&&& \\
{\it Radial-velocities (joint analysis)}\\
$A_{2}$ [\ms] & $\mathcal{U}(0,100)$ & $5.6^{_{+2.9}}_{^{-1.5}}$ & $5.7^{_{+2.7}}_{^{-1.6}}$ \\
$\lambda_{1}$ [d] & $\mathcal{U}(3.0,500)$ & $26^{_{+27}}_{^{-16}}$ & $37^{_{+11}}_{^{-19}}$ \\
$\lambda_{2}$ & $\mathcal{U}(0, 3)$ & $1.88\pm0.79$ & $1.72\pm0.88$ \\
$P_{rot}$ [d] & $\mathcal{N}(24.73,2.2)$ & $25.5\pm1.7$ & $25.8^{_{+1.2}}_{^{-1.8}}$ \\
&&& \\
{\it Photometry (learning on the light curve for activity filtering)}\\
$A_{1}$ & $\mathcal{U}(0,1)$ & \multicolumn{2}{c}{$0.00293\pm0.00032$} \\
$l$ [d] & $\mathcal{U}(0,5)$ & \multicolumn{2}{c}{$1.444\pm0.056$} \\
&&& \\
{\it Photometry (period learning on the light curve)}\\
$A_{2}$ & $\mathcal{U}(0,100)$ & \multicolumn{2}{c}{$0.00315\pm0.00044$} \\
$\lambda_{1}$ [d] & $\mathcal{U}(3.0,500)$ & \multicolumn{2}{c}{$21.1^{_{+3.6}}_{^{-5.9}}$}  \\
$\lambda_{2}$ & $\mathcal{U}(0, 3)$ & \multicolumn{2}{c}{$0.411^{_{+0.057}}_{^{-0.027}}$} \\
$P_{rot}$ [d] & $\mathcal{U}(15,40)$ & \multicolumn{2}{c}{$24.73^{_{+0.30}}_{^{-2.20}}$} \\
&&& \\
{\it Full Width Half Maximum (period learning on the FWHM time series)}\\
$A_{2}$ [\kms] & $\mathcal{U}(0,10)$ & \multicolumn{2}{c}{$0.01720^{_{+0.0051}}_{^{-0.0028}}$} \\
$\lambda_{1}$ [d] & $\mathcal{U}(0,1000)$ & \multicolumn{2}{c}{$3.9^{_{+50.0}}_{^{-1.4}}$}  \\
$\lambda_{2}$ & $\mathcal{U}(0,1000)$ & \multicolumn{2}{c}{$118^{_{+600}}_{^{-120}}$} \\
$P_{rot}$ [d] & $\mathcal{U}(5,40)$ & \multicolumn{2}{c}{$25.8^{_{+7.7}}_{^{-12.0}}$} \\
&&& \\
{\it Bisector Inverse Slope (period learning on the BIS time series)}\\
$A_{2}$ [\kms] & $\mathcal{U}(0,10)$ & \multicolumn{2}{c}{$0.042^{_{+0.150}}_{^{-0.030}}$} \\
$\lambda_{1}$ [d] & $\mathcal{U}(0,1000)$ & \multicolumn{2}{c}{$764^{_{+170}}_{^{-280}}$}  \\
$\lambda_{2}$ & $\mathcal{U}(0,1000)$ & \multicolumn{2}{c}{$453\pm380$} \\
$P_{rot}$ [d] & $\mathcal{U}(5,40)$ & \multicolumn{2}{c}{$22\pm13$} \\
&&& \\
{\it H$_{\alpha}$ index (period learning on the H$_{\alpha}$ index time series)}\\
$A_{2}$ & $\mathcal{U}(0,10)$ & \multicolumn{2}{c}{$0.36^{_{+0.88}}_{^{-0.30}}$} \\
$\lambda_{1}$ [d] & $\mathcal{U}(0,1000)$ & \multicolumn{2}{c}{$737\pm260$}  \\
$\lambda_{2}$ & $\mathcal{U}(0,1000)$ & \multicolumn{2}{c}{$281^{_{+440}}_{^{-230}}$} \\
$P_{rot}$ [d] & $\mathcal{U}(5,40)$ & \multicolumn{2}{c}{$26.3^{_{+5.7}}_{^{-2.8}}$} \\
&&& \\
{\it Na~{\sc i} D index (period learning on the Na~{\sc i} D index time series)}\\
$A_{2}$ & $\mathcal{U}(0,10)$ & \multicolumn{2}{c}{$0.01500^{_{+0.0052}}_{^{-0.0031}}$} \\
$\lambda_{1}$ [d] & $\mathcal{U}(0,1000)$ & \multicolumn{2}{c}{$10.8\pm2.7$}  \\
$\lambda_{2}$ & $\mathcal{U}(0,1000)$ & \multicolumn{2}{c}{$435\pm390$} \\
$P_{rot}$ [d] & $\mathcal{U}(5,40)$ & \multicolumn{2}{c}{$22\pm12$} \\
&&& \\
{\it S$_{\rm MW}$ index (period learning on the S$_{\rm MW}$ index time series)}\\
$A_{2}$ & $\mathcal{U}(0,10)$ & \multicolumn{2}{c}{$0.0304^{_{+0.0090}}_{^{-0.0061}}$} \\
$\lambda_{1}$ [d] & $\mathcal{U}(0,1000)$ & \multicolumn{2}{c}{$7.7\pm1.7$}  \\
$\lambda_{2}$ & $\mathcal{U}(0,1000)$ & \multicolumn{2}{c}{$482\pm360$} \\
$P_{rot}$ [d] & $\mathcal{U}(5,40)$ & \multicolumn{2}{c}{$23\pm12$} \\
&&& \\

\end{longtable}
\end{landscape}
\end{appendix}

\end{document}